\documentstyle[preprint,aps,eqsecnum]{revtex}   

\newcommand{\beq}{\begin{equation}}
\newcommand{\eeq}{\end{equation}}
\newcommand{\beqa}{\begin{eqnarray}}
\newcommand{\eeqa}{\end{eqnarray}}
\newcommand{\bsnn}{\mbox{$B \to X_s\,\nu\,\bar\nu$}}
\newcommand{\bsntnt}{\mbox{$B \to X_s\,\nu_\tau\,\bar\nu_\tau$}}
\newcommand{\bqnn}{\mbox{$B \to X_q\,\nu\,\bar\nu$}}
\newcommand{\bsmm}{\mbox{$B \to X_s\,\mu^+\,\mu^-$}}
\newcommand{\bsee}{\mbox{$B \to X_s\,e^+\,e^-$}}
\newcommand{\bsll}{\mbox{$B \to X_s\,\ell^+\,\ell^-$}}
\newcommand{\bksee}{\mbox{$B \to K^*\,e^+\,e^-$}}

\newcommand{\bsg}{\mbox{$B \to X_s\,\gamma$}}
\newcommand{\zbb}{\mbox{$Z \to b\,\bar b$}}
\newcommand{\btn}{\mbox{$B \to \tau\,\bar\nu$}}
\newcommand{\bctn}{\mbox{$B \to X_c\,\tau\,\bar\nu$}}
\newcommand{\bsz}{\mbox{$bsZ$}}
\newcommand{\bszp}{\mbox{$bsZ^\prime$}}
\newcommand{\BR}{{\rm BR}}
\newcommand{\Rbs}{\mbox{${\scriptstyle \not \! \! \; R}$}}
\newcommand{\cttp}{\mbox{$\cot\theta^\prime$}}
\newcommand{\cttps}{\mbox{$\cot^2\theta^\prime$}}

\newcommand{\cttpf}{\mbox{$\cot^4\theta^\prime$}}
\newcommand{\ct}{\mbox{$\widetilde C$}}

\def\overarrow#1{\ \ \rlap{\hbox{$\longrightarrow$}}
                        {\raise 6pt \hbox{\ $\> \scriptstyle#1$}}\quad\ }

\def\npb#1{Nucl.\ Phys.\ {\bf B #1}}
\def\plb#1{Phys.\ Lett.\ {\bf B #1}}
\def\prd#1{Phys.\ Rev.\ {\bf D #1}}
\def\prl#1{Phys.\ Rev.\ Lett. {\bf #1}}

\def\zpc#1{Z.~Phys.\ {\bf C #1}}
\def\prep#1{Phys.\ Rep.\ {\bf #1}}

\def\ijmpa#1{Int.\ J.\ Mod.\ Phys.\ {\bf A #1}}

\def\ea{{\it et al.}}
\def\ib{{\it ibid.}}
\def\eg{{\it e.g.}}

\def\Lra{\Longrightarrow}

\begin{document}

\draft

{\tighten
\preprint{\vbox{\hbox{WIS-95/49/Oct-PH}
                \hbox{CALT-68-2022}
                \hbox{hep-ph/9510378} }}

\title{First limit on inclusive \bsnn\ decay \\ and constraints on new physics}

\author{Yuval Grossman\,$^a$, Zoltan Ligeti\,$^b$ and Enrico Nardi\,$^a$}

\footnotetext{\footnotesize E-mail: ftyuval@weizmann.weizmann.ac.il,
zoltan@theory.caltech.edu, ftnardi@wicc.weizmann.ac.il}

\address{ \vbox{\vskip 0.truecm}
  $^a$Department of Particle Physics \\
  Weizmann Institute of Science, Rehovot 76100, Israel \\
\vbox{\vskip 0.truecm}
  $^b$California Institute of Technology, Pasadena, CA 91125 }

\maketitle

\begin{abstract}%
The inclusive \bsnn\ decay rate, on which no experimental bound exists to date,
can be constrained by searching for large missing energy events in $B$ decays.
Carefully examining the experimental and theoretical aspects of such an
analysis, we argue that the  published ALEPH limit on $\BR(\btn)$ implies,
conservatively, the bound $\BR(\bsnn)<3.9\times10^{-4}$, which is less than one
order of magnitude above the standard model prediction.  The LEP collaborations
could significantly improve this bound by a dedicated experimental analysis.
We study the constraints this new limit imposes on various extensions of the
standard model.  We derive new bounds on the couplings of third generation
fermions in models with leptoquarks, and in supersymmetric models without
R-parity. We also constrain models where new gauge bosons are coupled
dominantly to the third generation, such as TopColor models and models based on
horizontal gauge symmetries. For models which predict an enhanced effective
$\bsz$ vertex, the constraint from \bsnn\ is competitive with the limits from
inclusive and exclusive \bsll\ decays.

\end{abstract}

} 

\newpage

\section{Introduction}

Recent progress in experiment and theory has made flavor changing neutral
current (FCNC) $B$ decays a stringent test of the Standard Model (SM) and a
powerful probe of New Physics (NP).  The CLEO Collaboration observed the
exclusive decay $B\to K^*\,\gamma$ \cite{CLEO-bsg-ex} as well as the inclusive
decay $B\to X_s\,\gamma$ \cite{CLEO-bsg-in}.  The UA1 upper limit on the
inclusive decay \bsmm\ \cite{UA1}, and the recent CLEO \cite{CLEO-bKll} and CDF
\cite{CDF-bKll} upper limits on the exclusive decays
$B\to K^{(*)}\,\ell^+\,\ell^-$, are less than one order of magnitude above the
SM predictions.  These decays are likely to be observed within the next few
years.  These experimental results constitute a set of very strong constraints
on several possible sources of NP.

The FCNC decay \bsnn\ is also very sensitive to extensions of the SM, and
provides a unique source of constraints on some NP scenarios which predict a
large enhancement of this decay mode.  In particular, the \bsntnt\ mode is very
sensitive to the relatively unexplored couplings of third generation fermions.
However, no experimental upper bound on this decay mode has been
established to date.

The decay \bsnn\ can be searched for through the large missing energy
associated with the two neutrinos.  Using such techniques, the ALEPH
\cite{ALEPHfirst,ALEPH}, L3 \cite{L3}, and OPAL \cite{OPAL} collaborations have
been able to measure the inclusive \bctn\ decay rate \cite{FLNN}.  The large
missing energy in this case is associated with  the two neutrinos in the decay
chain \bctn\ followed by $\tau\to\nu\,X$.  From the absence of excess events
with very large missing energy, ALEPH also established the
90\%CL bound \cite{ALEPH}
\beq \label{btnbound}
  \BR(\btn) < 1.8 \times 10^{-3} \,.
\eeq

In this paper we point out that a similar analysis of the same data also 
implies a limit on $\BR(\bsnn)$. While a detailed and complete analysis 
of this set of data can only be performed by the ALEPH Collaboration, 
we show that using some conservative and
simplifying assumptions it is possible to derive the bound
\beq \label{newlimit}
  \BR(\bsnn) < 3.9 \times 10^{-4} \,,
\eeq
which is less than one order of magnitude above the SM prediction \cite{BuBu}
\beq \label{smbr}
  \BR^{\rm SM}(\bsnn) \approx 5 \times 10^{-5} \,.
\eeq
We expect that a dedicated analysis of the LEP collaborations will strengthen
our bound.  If background subtraction can be performed, or extra experimental
cuts can reduce the background, then we estimate that using the full LEP--I
data sample a 90\%CL bound of order
\beq \label{estbr}
  \BR(\bsnn) < (1-2) \times 10^{-4} \,,
\eeq
could be within the reach of the LEP experiments.

In section II we discuss, in a model independent manner, the inclusive \bsnn\
decay rate and the missing energy spectrum.  In section III we describe the
theoretical issues involved in relating the limits on large missing energy
events to the \bsnn\ decay rate, and we estimate the corresponding theoretical
uncertainties.  This section also contains the details of the derivation of our
bound (\ref{newlimit}).  While the limit we obtain does not allow a direct test
of the SM, it still implies stringent constraints on various new physics
scenarios.  In section IV we analyze the implications of existing measurements
of FCNC processes, particularly \bsg\ and \bsmm, for \bsnn, in various classes
of NP models.  Section V contains an extensive discussion of the constraints
implied by the bound (\ref{newlimit}) on various extensions of the SM.  We
derive new constraints on models in which the couplings of the third family
fermions differ from those of the first two generations, and we present
numerical limits on several NP parameters.  Finally, section VI contains a
summary of our results and the conclusions.

\section{The \bsnn\ decay rate}

{}From the theoretical point of view, the decay \bsnn\ is a very clean process,
since both the perturbative $\alpha_s$ and the non-perturbative $1/m_b^2$
corrections are known to be small.
Furthermore, in contrast to the decay \bsll, which suffers from background such
as $B\to X_s\,J/\psi\to X_s\,\ell^+\,\ell^-$, there are no analogous
long-distance QCD contributions, since there are no intermediate states that
can decay into a neutrino pair.  Therefore, the decay \bsnn\ is well suited to
search for and constrain NP effects.

As our aim is to derive constraints on NP scenarios, we discuss the missing
energy spectrum in a model independent framework.  Limits on NP parameters can
then be derived by comparing the experimental bound with the theoretical
expressions as derived in specific models.  A model independent expression for
the missing energy spectrum can be straightforwardly obtained from the general
result for muon decay \cite{book}.  Under the only assumption of two component
left-handed neutrinos (possible neutrino mass effects are at most
of order $m_\nu/m_b <10^{-2}$), the most general form of the four-fermion
interaction responsible for $B \to X_q\,\nu_i\,\bar\nu_j$ reads
\beq \label{Lgeneral}
  {\cal L} = C_L\, O_L + C_R\, O_R \,,
\eeq
where
\beq \label{OL-OR}
O_L = [\bar q_L \, \gamma_\mu\, b_L]\,
  [\bar \nu^i_L\, \gamma^\mu \nu^j_L] \,, \qquad
O_R = [\bar q_R \, \gamma_\mu\, b_R]\,
  [\bar \nu^i_L\, \gamma^\mu \nu^j_L]\,.
\eeq
Here $L$ and $R$ denote left- and right-handed components, $q=d,s$, and
$i,j=e,\mu,\tau$.  In this article we adopt the notation that a generic
$B$ meson contains a $b$ quark, rather than a $\bar b$ quark.
As the flavors of the decay products are not detected, in certain
models more than one final state can contribute to the observed decay rate.
Then, in principle, both $C_L$ and $C_R$ carry three indices $q,\,i,\,j$,
which label the quark and neutrino flavors in the final state.
Throughout our discussion, we shall only keep track of these indices when they
are important, otherwise we will suppress them.

At lowest order, the missing energy spectrum in the $B$
rest-frame is given by \cite{book}
\beq \label{shape}
{{\rm d}\Gamma(B \to X_q\,\nu_i\,\bar\nu_j)\over{\rm d}x} =
{m_b^5\over96\pi^3}\,
  \left(|C_L|^2+|C_R|^2\right)\, {\cal S}(r,x) \,.
\eeq
Here we have not yet summed over the neutrino flavors.
The function
${\cal S}(r,x)$ describes the shape of the missing energy spectrum
\beq \label{shapefn}
{\cal S}(r,x) = \sqrt{(1-x)^2-r}\,
  \Big[ (1-x)\,(4x-1) + r\,(1-3x) - 6\eta\sqrt{r}\,(1-2x-r) \Big] \,.
\eeq
The dimensionless variable $x={E_{\rm miss}/m_b}$ can range between
$(1-r)/2\leq x\leq1-\sqrt{r}$, and $r={m_s^2/m_b^2}$.  The parameter
$\eta=-{\rm Re}(C_L\,C_R^*)/(|C_L|^2+|C_R|^2)$, which is the analog of the
Michel parameter in $\mu$-decays, ranges between
$-\frac12\leq\eta\leq\frac12$.

In the SM, \bsnn\ proceeds via $W$ box and $Z$ penguin diagrams, therefore
only $O_L$ is present and $\eta=0$.  The corresponding coefficient reads
\cite{inami-lim}
\beq \label{HeffSM}
C_L^{\rm SM} = {\sqrt2\,G_F\,\alpha\over \pi \sin^2\theta_W}\,
  V_{tb}^*\,V_{ts}\, X_0(x_t) \,,
\eeq
where $x_t=m_t^2/m_W^2$, and
\beq \label{eqX}
  X_0(x) = {x\over8}\, \left[{2+x\over x-1}+{3x-6\over(x-1)^2}\,\ln x \right].
\eeq
In the limit of large top quark mass, $X_0$ has a quadratic dependence on
$m_t$, $X_0(x_t)\sim x_t/8$.  Therefore, the main source of uncertainty in the
SM prediction for the total decay rate comes from the uncertainty in $m_t$.

The leading $1/m_b^2$ and $\alpha_s$ corrections to the SM result (calculated
in the free quark decay model) are known.  The $\alpha_s$ correction to the
total decay rate is given by replacing $X_0(x_t)$ in Eq.~(\ref{HeffSM}) by
\cite{BuBu}
\beq
X_0(x_t) \to \bigg[ X_0(x_t) + {\alpha_s\over4\pi}\,X_1(x_t) \bigg]
  \bigg[ 1 - {\alpha_s\over3\pi}\,\bigg(\pi^2-{25\over 4}\bigg) \bigg] \,.
\eeq
Here the second term represents the correction to the matrix element of $O_L$.
This term cancels almost completely when the \bsnn\ branching ratio is
normalized to the semileptonic $B\to X_c\,e\,\bar\nu$ rate (see
Eq.~(\ref{BRNP}) below).  The first term contains the QCD correction to the box
and penguin diagrams.  We do not display here the explicit form of $X_1(x_t)$,
which can be found in \cite{BuBu}.  The most important effect of this
correction is to reduce the scale dependence of the SM prediction from about
$\pm10\%$ to below $\pm2\%$ \cite{BuBu}.

The $1/m_b^2$ correction to the contribution of the $O_L$ operator to the
missing energy spectrum can be read off from \cite{Adametal}.  The result is
the following modification of the function ${\cal S}(r,x)$ in
Eq.~(\ref{shapefn})
\begin{eqnarray} \label{mb2corr}
&& {\cal S}(r,x) \to {\cal S}(r,x) \nonumber\\*
&& + {1\over\sqrt{(1-x)^2-r}}\, \bigg\{
  {\lambda_1\over6m_b^2}\, [2(1-x)^2\,(19-38x+10x^2) -r(61-23r-122x+52x^2)]
  \nonumber\\*
&& + {\lambda_2\over2m_b^2}\, [(1-x)(47-126x+96x^2-20x^3) -
  r(70-23r-125x+52x^2)] \bigg\} \nonumber\\
&& + \bigg[{\lambda_1\over48m_b^2}\,(5-r)-{\lambda_2\over16m_b^2}\,(1-5r)\bigg]
  \, (1-r)^3\, \delta\bigg(\frac12(1-r)-x\bigg) \nonumber\\*
&& + {\lambda_1\over96m_b^2}\, (1-r)^5\, \delta'\bigg(\frac12(1-r)-x\bigg)\,.
\end{eqnarray}
Here $\lambda_1$ and $\lambda_2$ are related to the kinetic energy of the
$b$ quark inside the $B$ meson and to the mass splitting between the
$B$ and the $B^*$ mesons, respectively.
Experimentally, $\lambda_2\approx 0.12\,{\rm GeV}^2$, and following the
discussion in \cite{gl}, we use $0<-\lambda_1< 0.5\,{\rm GeV}^2$.
When integrated over the spectrum, this correction amounts to about 3\%
suppression of the total decay rate.  Even this small correction
cancels almost entirely when the \bsnn\ branching ratio is normalized to
the semileptonic $B\to X_c\,e\,\bar\nu$ rate (see Eq.~(\ref{BRNP}) below).
As we will discuss in the next section, although the \bsnn\ branching
fraction in the SM is known rather precisely, the theoretical prediction
for the missing energy spectrum is more uncertain.

In several NP models $O_R$ is also present.  The structure of the operator
product expansion \cite{CGG,MaWi,BKSV} shows that to all orders in the
$\alpha_s$ and $1/m_b$ expansions, both the perturbative and non-perturbative
corrections to the contribution of $O_R$ to the missing energy spectrum are
identical to those of $O_L$.  This holds as long as the phase space is
symmetric in the two leptons (this can be violated only by negligible neutrino
mass effects).  Thus, the shape of the missing energy spectrum is model
independent, up to possible small interference effects between $O_L$ and $O_R$,
of order $m_s/m_b$.  Once $C_L$ and $C_R$ are computed in any particular model,
the unknown contributions to the total decay rate are only
${\cal O}(\alpha_s^2;\, \alpha_s\,m_s/m_b;\, \alpha_s\,\Lambda^2/m_b^2;\,
\Lambda^3/m_b^3;\, m_s\,\Lambda^2/m_b^3)$, where $\Lambda$
denotes some scale of order $\Lambda_{\rm QCD}$.

In some NP models $O_L$ and $O_R$ are simultaneously present with coefficients
of comparable size, giving rise to interference effects proportional to $\eta$,
which modify the shape of the missing energy spectrum.  Due to the $m_s/m_b$
suppression, this effect is always small, except close to the endpoint region
($x\sim1-\sqrt r$), where the leading term in (\ref{shapefn}) also becomes of
order $\sqrt{r}$.  Therefore, close to the endpoint, the corrections to the
shape of the spectrum could be relevant.  Thus, besides the quark mass ratio,
an additional uncertainty in (\ref{shape}) is associated with the value of
$\eta$.  The softest missing energy spectrum, and thus the most conservative
bound on the \bsnn\ branching ratio, is obtained by using a large quark mass
ratio $r\simeq0.002$ (corresponding to $m_s\simeq0.2\,$GeV and
$m_b\simeq4.8\,$GeV), and $\eta=-\frac12$.  The bound on the branching fraction
derived using these values of $r$ and $\eta$ holds in any model.

In order to find the constraints on NP, we will express the total decay rate in
terms of two ``effective" coefficients $\ct_L$ and $\ct_R$, which can be
computed in terms of the parameters of any NP model and are directly related to
the experimental measurement (see (\ref{npbound})).
To remove the large uncertainty in the total decay rate associated with the
$m_b^5$ factor, it is convenient to normalize $\BR(\bsnn)$ to the semileptonic
rate $\BR(B\to X_c\,e\,\bar\nu)$, since the experimental value of the latter is
known quite precisely.  The contribution from $B\to X_u\,e\,\bar\nu$, as well
as possible NP effects on the semileptonic decay rate are negligible.  In
constraining NP, we can also set $m_s=0$ and neglect both order $\alpha_s$ and
$1/m_b^2$ corrections.  This is justified, since when averaged over the
spectrum these effects are very small, and would affect the numerical bounds on
the NP parameters only in a negligible way.
For the total $B \to X_q\,\nu_i\,\bar\nu_j$ decay rate
into all possible $q=d,s$ and $i,j=e\,,\mu\,,\tau$
final state flavors, we then obtain
\beq \label{BRNP}
{\BR(B \to X \,\nu\,\bar\nu)
\over \BR(B\to X_c\,e\,\bar\nu)} =
  {\ct^2_L + \ct^2_R \over|V_{cb}|^2\, f_{PS}(m_c^2/m_b^2)}\,,
\eeq
where $f_{PS}(x)=1-8x+8x^3-x^4-12x^2\ln x$ is the usual phase-space factor,
and we defined
\beq \label{ctLR}
\ct^2_L = {1\over 8 G_F^2}\, \sum_{q,i,j} {\left|C_L^{qij}\right|}^2 , \qquad
\ct^2_R = {1\over 8 G_F^2}\, \sum_{q,i,j} {\left|C_R^{qij}\right|}^2 .
\eeq

The SM prediction for the branching ratio in Eq.~(\ref{smbr}) is obtained by
inserting into Eq.~(\ref{BRNP}) the semileptonic rate
${\rm BR}(B\to X_c\,e\,\bar\nu)\approx10.5\%$ \cite{PDG},
$f_{PS}(m_c^2/m_b^2)\approx0.5$, together with
\beq \label{ctSM}
  \Big( \ct_L^{\rm SM} \Big)^2 \approx 3.8 \times 10^{-7} \,,
\eeq
which follows from (\ref{HeffSM}) by using
$|V_{tb}^*\,V_{ts}|/|V_{cb}|\approx1$, $|V_{cb}|\approx0.04$,
$\alpha(m_Z)\approx1/129$, $\sin^2\theta_W\approx0.23$,
and $m_t\approx180\,$GeV \cite{PDG}.
In a SM analysis it would be more natural to factor out from the definitions of
$\ct_{L,R}$ the small mixing angles $V_{tb}^*\,V_{ts}$ as well as $\alpha$,
resulting in a dimensionless coefficient of order unity.  However, there is no
reason to do so for $\ct_R$, and since in several NP models $\ct_{L,R}$ are
induced at the tree level, they do not contain any small parameters analogous
to $\alpha$ or to the CKM angles. In fact, often the only suppression factors
in these coefficients come from inverse powers of some large mass scale.

Comparing the limit (\ref{newlimit}) with (\ref{BRNP}) yields the following
constraint on possible NP contributions:
\beq \label{npbound}
\ct_L^2+\ct_R^2 < 3.0 \times 10^{-6}
  \left[{\BR(\bsnn) \over 3.9 \times 10^{-4}}\right].
\eeq
In section V we will constrain NP models by comparing this limit with
the various theoretical predictions for the coefficients $\ct_{L,R}$.
We will quote the bounds that can be derived on each NP parameter, even
in those cases when $O_L$ and $O_R$ are simultaneously present and
certain combinations of the NP parameters may be better constrained.

\section{Experimental analysis}

In this section we discuss the theoretical uncertainties involved in the
experimental analysis at LEP, and we derive the bound (\ref{newlimit}).  To
obtain a limit on \bsnn\ using the data from ALEPH \cite{ALEPH}, we have to
estimate the relative efficiency of \bsnn\ and \btn\ to pass the experimental
missing energy cut (ALEPH used the cut $E_{\rm miss}>35\,$GeV).  We use the
\bsnn\ missing energy spectrum as given in Eqs.~(\ref{shape}) and
(\ref{shapefn}), together with the conservative values $r=0.002$ and
$\eta=-\frac12$ which, as discussed in the previous section, lead to the
softest missing energy spectrum.  As a result, the bound we derive holds 
in any model.

\subsection{Theoretical uncertainties}

The largest theoretical uncertainty is related to the reliability of the
theoretical calculation of the missing energy spectrum near its endpoint
($x\sim1-\sqrt r$).  While both the $\alpha_s$ and $1/m_b^2$ corrections
to the total decay rate are small, this is not true point-by-point for the
differential energy spectrum.

The $1/m_b$ expansion for inclusive semileptonic decays of hadrons containing a
single $b$ quark based on an operator product expansion \cite{CGG} is not
reliable when the invariant mass of the decay product hadronic system is small.
In the endpoint region, near $x=1-\sqrt{r}$, there are large corrections to the
free quark decay prediction, and the spectrum has to be smeared to get a
reliable result.  This smearing region has to be chosen large enough so that
after smearing, the $1/m_b^2$ corrections in (\ref{mb2corr}) produce only small
effects on the decay spectrum \cite{MaWi}.  Following this criterion, we find
that in the present case the smearing region extends about $0.5-0.7\,$GeV from
the maximal value of the missing energy.

The order $\alpha_s$ corrections to the missing energy spectrum can be read off
from \cite{Kuhn}, and are less problematic.  While there are large logarithms
for small values of the hadronic invariant mass, the perturbative correction to
the hadron energy spectrum is smooth in the limit of vanishing hadronic energy.
Thus, the perturbative corrections to the missing energy spectrum are small and
reliably calculable in the region relevant for our analysis
\cite{Kuhn,Adametal}.

The large non-perturbative corrections near maximal missing energy could
represent a problem for the analysis of \bsnn, since the experimental cuts
select events with large missing energy, thus enhancing the weight of the
problematic region.  However, the experimental search is at the endpoint region
in the laboratory frame.  Events with large missing energy in the laboratory
frame are not necessarily those with missing energy close to the endpoint in
the $B$ rest-frame.  We need to take into account the large boost from the $B$
rest-frame to the laboratory frame.  We define
$P_{\epsilon[{\rm GeV}]}(E_{\rm cut}[{\rm GeV}])$
as the fraction of the events with $E_{\rm miss}>E_{\rm cut}$ in the laboratory
frame, that come from the missing energy region in the $B$ rest-frame between
the endpoint and $E_0=m_b-m_s-\epsilon$.  To simulate the $B$ hadron energy in
the laboratory frame, we use in our Monte Carlo code the Peterson fragmentation
function \cite{pet}, and $m_b\approx4.8\,$GeV.  We find the following
representative numbers
\beqa \label{dept}
&&  P_{0.5}(40) \approx 0.12\,,  \qquad  P_{0.5}(35) \approx 0.11\,,  \qquad
  P_{0.5}(30) \approx 0.09\,, \nonumber \\*
&&  P_{0.7}(40) \approx 0.21\,,  \qquad  P_{0.7}(35) \approx 0.19\,,  \qquad
  P_{0.7}(30) \approx 0.17\,.
\eeqa
The results in (\ref{dept}) show some sensitivity to the size of the smearing
region, while they suggest that varying the experimental cut has a much
smaller effect.  We have also checked that these figures are not very sensitive
to the choice of the quark masses and of $\eta$.

We conclude that the missing energy spectrum in the laboratory frame of the LEP
experiments can be estimated reliably from the spectator model.  However, the
uncertainty related to the endpoint region of the missing energy spectrum in
the $B$ rest-frame should be taken into account.  To estimate this uncertainty,
we rely on the fact that the size of the endpoint region is chosen such that
the integration over it can be trusted.  Assigning to the fraction of the
events coming from endpoint region the lowest possible value of the missing
energy, $m_b-m_s-\epsilon$, we can bound the uncertainty related to using the
spectator model.  For $\epsilon=0.7\,$GeV we find  in this way a 5\% reduction
in the total number of events that pass the $35\,$GeV cut.  We consider this as
a reliable estimate of the theoretical uncertainty of the analysis related to
the use of the spectator model.

A final remark is in order.  At LEP, $b$ quarks hadronize into a variety of $b$
hadrons.  However, by the time it decays, the $b$ quark is contained either in
a $B$ meson or in a $\Lambda_b$ baryon.  The fraction of $\Lambda_b$ baryons at
LEP has been measured to be about 10\% \cite{ALEPHbrate}.  In the limit when
the $b$ quark is treated as very heavy, the missing energy spectrum from $B$
decays should be similar to that from $\Lambda_b$ decays, up to small effects
originating from the polarization of the baryons. A possible indication for
significant heavy quark symmetry violating effects, is the experimentally
measured lifetime ratio between the $B$ and the $\Lambda_b$ hadrons, which
appears to be larger than the theoretical prediction (for a recent discussion
see, {\it e.g.}, \cite{bigirev}).  However, the resolution of this problem is
most likely related either to experimental issues or to the theoretical
calculation of the hadronic decay widths, so it is probably irrelevant for the
analysis in this paper.  The polarization of the $\Lambda_b$ baryons produced
at LEP has been measured to be $-30\pm30$\% \cite{ALEPHpol}.  A non-vanishing
left-handed polarization enhances the missing energy from $\Lambda_b$ decays,
and therefore neglecting it is conservative.

\subsection{Derivation of the bound}

After this discussion, we are ready to estimate the bound on $\BR(\bsnn)$.
To translate the ALEPH bound on \btn\ \cite{ALEPH} into a limit on \bsnn,
we need to compare the theoretical predictions for the fraction of events that
pass the missing energy cut for the two decay modes.
The resulting bound on \bsnn\ is
stronger than that on \btn\ for the following reasons:
\vspace{-8pt}
\begin{itemize}\itemsep=-4pt
\item[($i$)]
The \btn\ decay is allowed only for the charged $B$ mesons,
while all $b$ flavored hadrons can decay through the parton level
process $b\to s\,\nu\,\bar\nu$;
\item[($ii$)]
In order to reject background from semileptonic $B$ decays,
only the hadronic $\tau$ decays were used in the ALEPH search for
\btn.\footnote{We thank Ian Tomalin for pointing this out to us.}
\item[($iii$)]
The missing energy spectrum is somewhat harder in \bsnn\ than in \btn\
decays, increasing the efficiency of the analysis.
\end{itemize} \vspace{-8pt}

To evaluate the last factor ($iii$), we need to estimate the missing energy
spectrum for the \btn\ decay as well.  In estimating the $\tau\to \nu\,X$
missing energy spectrum, we consider only two-body (20\%) and three-body (80\%)
hadronic $\tau$ decays.  Since a significant fraction of hadronic tau decays
are four- and five-body decays, the resulting missing energy spectrum for \btn\
is harder than the real one.  Therefore we get a conservative upper bound on
the \bsnn\ branching ratio, {\it i.e.}, weaker than what a detailed analysis
would obtain.  We take into account that the $\tau$ polarization decreases the
efficiency of the \btn\ analysis by about 20\%, as estimated in the ALEPH
analysis \cite{ALEPH}.  We find that in our approximation the efficiency of the
\bsnn\ decay to pass the ALEPH cut $E_{\rm miss}>35\,$GeV is about 15\% larger
than that of $\btn$ followed by hadronic $\tau$ decay.

Collecting the results of our previous discussion, we can finally
estimate that the bound on $\BR(\bsnn)$ is stronger then the ALEPH bound
on $\BR(\btn)$ by an overall factor
\beq \label{ratio}
R = R_{B^\pm}\, R_{\rm hadr}\, R_{\rm eff} \approx 0.21\,.
\eeq
Here $R_{B^\pm}\approx0.37$ is the ratio of $B^\pm$ mesons to the total
number of $b$ hadrons at ALEPH \cite{ALEPHbrate} and $R_{\rm hadr}\approx0.65$
is the hadronic $\tau$ branching fraction \cite{PDG}.
The factor $R_{\rm eff}$ accounts for the efficiency of the missing energy cut
in the $\btn$ decay followed by hadronic $\tau$ decays,
relative to that in \bsnn\ decays.  We conservatively use the upper bound
$R_{\rm eff}=0.90$ which includes the 5\% theoretical uncertainty related to
the reliability of the spectator model spectrum.  We expect the uncertainty
related to the use of the Peterson fragmentation function to cancel to a large
extent from the estimate of the relative efficiency of the \bsnn\ and \btn\
analyses.  Using (\ref{ratio}) and the experimental bound (\ref{btnbound}), we
find the limit given in Eq.~(\ref{newlimit}).

We would like to emphasize that the derivation of our bound relies on a set of
conservative simplifying assumptions, which could be avoided in a dedicated
experimental analysis.  Our limit (\ref{newlimit}) should be considered as a
conservative upper bound, to which we purposely do not (and cannot) assign a
confidence level.  We hope that a more complete and detailed investigation by
the LEP collaborations will be carried out.  The reward of such an analysis
could be a bound of order $(1-2)\times10^{-4}$, that is only a
factor $2-4$ above the SM prediction.

\section{Constraints on \bsnn\ from other processes}

Models that can give rise to large new contributions to the \bsnn\ decay often
predict an enhancement of other FCNC processes as well.  In this section we
analyze what constraints the existing experimental data on FCNC processes imply
for various NP models.  For each specific model, these constraints result in
upper limits on the allowed \bsnn\ decay rate.  Bounds on $\BR(\bsnn)$ can be
derived from the limits on rare processes, such as $K_L\to\mu^+\,\mu^-$,
$K\to\pi\,\nu\,\bar\nu$, $\epsilon_K$, $K-\bar K$ and $B-\bar B$ mixing.  The
most restrictive constraints are imposed by the measurements of the radiative
decay \bsg, and by the limits on inclusive and exclusive \bsll\ decays.

The radiative \bsg\ decay proceeds via photon penguin diagrams, and therefore
it is not directly related to \bsnn.  However, in many models the details of
the underlying physics imply relations between the $Z$ and the photon penguins.
In all these models the recent CLEO measurement \cite{CLEO-bsg-in}
\beq \label{bsginc}
  \BR(\bsg) = (2.32\pm 0.51\pm 0.32\pm 0.20)\times 10^{-4},
\eeq
which is in agreement with the SM, forbids large deviations from the SM
prediction for the \bsnn\ decay rate as well.

On the other hand, a large class of NP models predict (or can accommodate) an
enhanced \bsz\ effective vertex without giving rise to a large enhancement of
the $bs\gamma$ effective coupling.  Then the constraints from inclusive and
exclusive \bsll\ decays are important, as these decays, like \bsnn, are
dominated by $Z$ exchange.  In these models, a naive estimate of the ratio of
inclusive rates gives $\BR(\bsnn)/\BR(\bsll)\approx6$.  The factor of six
enhancement arises due to a factor of approximately two in the ratio between
the neutrino and the charged lepton couplings to the $Z$, and a factor of three
from the sum over the neutrino flavors.  A more precise calculation which
includes the photon exchange contribution to \bsll\ bounded by the CLEO
measurement of \bsg, as well as the sizable QCD corrections to \bsll\
\cite{GSaW,BuMu}, can increase the above ratio up to 7.  Hence, for this class
of models, the UA1 experimental limit on the inclusive \bsmm\ decay
\cite{UA1}\footnote{The UA1 experiment searched for events in the region
$3.9<E_{\mu\mu}<4.4\,$GeV.  However, the theoretical prediction for the
spectrum is uncertain in this endpoint region. Moreover, the theoretical
spectrum shown in Ref.~\cite{UA1} (and presumably used to relate the bound on
the number of events in the experimental window to the quoted bound on the
total branching fraction) seems to be in disagreement with that calculated in
\cite{GSaW}.  Despite these uncertainties we use the published bound.}
\beq \label{brinc}
  \BR(\bsmm) < 5 \times 10^{-5} \,,
\eeq
implies
\beq \label{bsnnUA1}
  \BR(\bsnn) \lesssim 3.5 \times 10^{-4} \quad \Lra \quad
  \ct_L^2+\ct_R^2 < 2.7 \times 10^{-6}\,.
\eeq
This limit is comparable with the limit (\ref{newlimit}), however it is weaker
than the expected LEP sensitivity (\ref{estbr}).  This underlines the
importance of a more detailed experimental analysis aimed at searching for
\bsnn, and shows that LEP measurements could compete with future new data from
CLEO and CDF in constraining models which predict an enhanced $bsZ$ coupling.
Moreover, as we have emphasized, the neutrino mode is particularly
interesting since it is theoretically cleaner than the charged lepton
modes: ($i$) there are no long-distance QCD effects; ($ii$) the
short-distance QCD corrections are small; ($iii$) there is no photon penguin
contribution, and therefore this process can be straightforwardly related to
the effective $bsZ$ vertex.  In conclusion, although the constraints
resulting  from our
limit (\ref{newlimit}) are numerically slightly weaker than those implied by
$\bsmm$ (\ref{bsnnUA1}), we think that they are more reliable.

The exclusive dilepton decays are also sensitive to an enhanced $bsZ$ vertex,
and provide additional constraints.  The exclusive decay modes
$B\to K^{(*)}\,\ell^+\,\ell^-$ have been searched for by the CLEO
\cite{CLEO-bKll} and the CDF \cite{CDF-bKll} collaborations. For our purposes,
the most restrictive limit has been established by CLEO \cite{CLEO-bKll}
\beq \label{cleo-bkee}
  \BR(\bksee) < 1.6 \times 10^{-5} \,.
\eeq
However, the interpretation of this limit is obscured by the significant
model dependence of the exclusive $B\to K^{(*)}$ form factors.
The estimates for the ratio
\beq
  \rho \equiv {\BR(B \to K^*\,e^+\,e^-) \over \BR(B\to X_s\,e^+\,e^-)} \,,
\eeq
range between $0.10$ and $0.35$ \cite{estR}.
Clearly, the value of $\rho$ is crucial for relating the limits on exclusive
decays to a limit on the FCNC transition at the quark level, for which the
SM predicts \cite{GSaW}\footnote{%
The spread in the SM predictions in the literature is in part due to
the different signs for the $O_7-O_9$ interference term in various papers.
(We use here the notation of Ref.~\cite{BuMu}.)}
\beq \label{bsll-th}
  \BR(\bsee) \approx 10 \times 10^{-6} \,, \qquad
  \BR(\bsmm) \approx 7 \times 10^{-6} \,.
\eeq
It is also questionable whether the above estimates of $\rho$, based on SM
computations, are reliable for the analysis of NP scenarios.  In fact, in the
SM a significant contribution to \bksee\ comes from soft photons, corresponding
to small $e^+ e^-$ invariant mass.  However, in the NP models under
consideration $Z$ penguins dominate, and therefore $\rho$ is expected to be
smaller than in the SM.

The only known model independent way to predict the $B\to K^{(*)}$ form factors
(besides lattice calculations) is to relate them to measurable semileptonic
decay form factors using heavy quark symmetries \cite{HQSrare}.  However, until
experimental information on the $B\to\rho\,\ell\,\bar\nu$ decay spectrum
becomes available, the dilepton spectrum can only be predicted in the small
window $4.0<m_{\ell^+\ell^-}<4.4\,$GeV from the $D\to K^*\,\ell^+\nu$ data.
(Even after the measurement of the $B\to\rho\,\ell\,\bar\nu$ spectrum, the
unknown symmetry breaking corrections will amount to an uncertainty of order
$20-30$\%.)

Rare $K$ decays also imply bounds on $Z$ penguins.  However, in relating them
to the \bsnn\ decay rate large uncertainties arise from poorly known quark
mixing angles. The only measured rate is
$K_L\to\mu^+\,\mu^-$, which receives large long-distance corrections.
Therefore, the short-distance parameters can only be extracted with large
uncertainties \cite{Ko}.  The decay $K^+\to\pi^+\,\nu\,\bar\nu$, however,
receives negligible long-distance contributions.  While the existing
experimental bound is about fifty times the SM prediction, an order of
magnitude improvement is expected from Brookhaven in the coming years.  In
summary, at present, for all the models in our analysis, rare $K$ decays are
less constraining than the rare $B$ decays discussed above.

Once these considerations are taken into account, it appears that the most
reliable constraints on an anomalous effective \bsz\ vertex, apart from \bsnn,
come from the limits on the inclusive \bsmm\ decay rate.  Numerically similar
bounds (or possibly slightly better -- depending on the adopted values of
model-dependent hadronic form factors) are provided by the bounds on exclusive
$B\to K^{(*)}\,\ell^+\,\ell^-$\ decay rates.

\section{New Physics}

Following the discussion in the previous section, it is useful to classify the
NP models which could enhance the SM prediction for \bsnn\ into three classes:
(A) Highly constrained models; (B) Weakly constrained models; (C) Unconstrained
models.

The first class (A) includes models in which the existing bounds on other FCNC
processes (mainly \bsg) imply that the rate for \bsnn\ cannot exceed the SM
prediction by any factor larger than two.  Thus, a bound on \bsnn\ of the order
of the LEP sensitivity (\ref{estbr}) does not imply any new constraint on the
underlying NP.  For completeness, we briefly explain below why the \bsnn\
branching fraction must be close to its SM value in the most popular models
belonging to this class: the minimal supersymmetric standard model; multi Higgs
doublet models; left-right symmetric models.

To class (B) belong all models which predict (or allow for) a large effective
\bsz\ (or \bszp) vertex.  We call these models ``weakly constrained", as the
limits on inclusive and exclusive \bsll\ already constrain them.  The limits
from (\ref{newlimit}) will not represent any numerical improvement over the
existing bounds on the parameters of these models.  However, our new limits are
more reliable, since the theoretical uncertainties involved in \bsnn\ are very
small.  Moreover, if a bound of the order (\ref{estbr}) can be obtained by the
LEP collaborations, then the constraints from \bsnn\ will become markedly
stronger than the present ones from \bsll.  A list of interesting NP models
belonging to this class includes: models with additional $Q=-\frac13$
isosinglet quarks; models with large \bszp\ vertex; models with an anomalous
effective $tcZ$ vertex; models with a heavy fourth generation; models with
anomalous $WWZ$ couplings; a class of extended technicolor models.  These
models  will be discussed in some detail below.

In the unconstrained models of class (C), the couplings responsible for
enhancing \bsnn\ are to a large extent independent of those constrained by any
other existing experimental bound.  Therefore, even a \bsnn\ decay rate orders
of magnitude above the SM prediction is still consistent with the existing
constraints, and the new limit (\ref{newlimit}) represents the most stringent
bound on the corresponding NP parameters.  Examples of theoretically
interesting models belonging to this class are: models with light leptoquarks;
supersymmetric models with broken R-parity; TopColor models; some models
based on non-Abelian horizontal gauge symmetries.

\subsection{Highly constrained models}

\subsubsection{Minimal Supersymmetric Standard Model (MSSM)}

In a large part of the SUSY parameter space, the MSSM (for review and
notations see, \eg, \cite{mssmrev}) is known to produce large effects on the
radiative decay \bsg, as well as on \bsll\ (see \cite{hhg,mssm-bbmr}).
The effects on \bsnn\ have been studied in \cite{mssm-bbmr,AGM}.  It was found
that the contributions to the rate can be non-negligible only for $\tan\beta$
close to unity, while for increasing values of $\tan\beta$ the prediction
rapidly converges to the SM value, regardless of the values of the other SUSY
parameters.  Even for $\tan\beta\simeq1$ it seems to be problematic to enhance
the rate up to the level observable at LEP, while keeping $\BR(\bsg)$ within
the experimental limits.  This is due to the fact that the SUSY contributions
to the $bs\gamma$ vertex tend to dominate over those to the \bsz\ vertex.  The
SUSY corrections to the \bsg\ decay can be kept small while allowing for a
large $\BR(\bsnn)$, only for low values of $\tan\beta$, and by invoking large
cancellations between the charged Higgs and chargino contributions.  It is then
conceivable that with a specific choice of several SUSY parameters, a
``fine-tuned" MSSM could produce $\BR(\bsnn)$ close to (\ref{estbr}).
However, we regard such a choice as unnatural.

\subsubsection{Multi Higgs Doublet Models (MHDM)}

MHDM (for review and notation see, \eg, \cite{hhg,Yuval}) are severely
constrained by \bsg, \zbb, \bctn, and lepton universality in tau decays. The
same is true for the more familiar two Higgs doublet models (2HDM), which
represent a subclass of the general MHDM with natural flavor conservation.  In
a general MHDM the single parameter $\tan\beta$ of the 2HDM is replaced by
three complex coupling constants, $X$, $Y$, and $Z$, which describe the Yukawa
interactions of the lightest charged scalar with the down-type quarks, up-type
quarks, and charged leptons, respectively.  The new $Z$ penguin diagrams
present in these models are related to new photon penguins, and thus are
severely constrained by \bsg\ and cannot contribute significantly to \bsnn.  A
large enhancement not affecting \bsg\ and \bsmm\ could arise for large enough
values of the $H^\pm\,\tau\,\nu_\tau$ Yukawa coupling via charged scalar box
diagrams involving two external $\nu_\tau$'s.  These box diagrams are
proportional to the combination $m_\tau^2\,Z^2\,(m_t^2\,Y^2+m_b\,m_s\,X^2)$
\cite{Yuval}.
However, $Y$ is constrained from \zbb, the coupling $Z$ from lepton
universality in tau decays, and the product $XZ$ from \bctn
\cite{Yuval,gl,ghn}, implying together
that also the contribution of the box diagrams to \bsnn\ has to be very small.

\subsubsection{Left--Right Symmetric Models (LRSM)}

In these models (for review and notations see, \eg, \cite{LRSrev}) new heavy
charged $W_R$ gauge bosons coupled to right-handed currents are exchanged in a
new set of electroweak penguin diagrams, which could give rise to deviations in
the rate for \bsnn.  However, new photon penguins would enhance \bsg\ as well.
The \bsg\ decay rate has been calculated in the minimal as well as in some
non-minimal versions of the LRSM.  Once the limits on $M_{W_R}$ and on the
$W_L-W_R$ mixing angle are imposed (from direct $W_R$ searches, $K_L-K_S$ mass
difference, {\it etc}.), the rate for \bsg\ cannot differ from the SM
prediction by more than about 50\% \cite{LRS-bsg}.  The CLEO measurement of
\bsg\ further constrains the parameter space of  non-minimal models to regions
where the $SU(2)_R$ gauge coupling is small.
In this region of parameters, the $Z$ penguins are
similarly suppressed, and therefore
significant deviations of the \bsnn\ decay rate from the SM prediction
are excluded.

\subsection{Weakly constrained models}

\subsubsection{Mixing of the $b$ with new exotic $Q=-\frac13$ quarks}

It is well known that the presence of new
heavy fermions with non-canonical $SU(2)$ transformations
($L$-handed singlets and/or $R$-handed doublets)
mixed with the standard leptons and quarks
would give rise to tree level FCNC in $Z$ interactions  \cite{fit}.
In particular, the presence of new $Q=-\frac13$ isosinglet quarks,
$D_L$ and $D_R$, as they appear for example in the {\bf 27} representation
of $E_6$ and in several superstring inspired extensions of the SM,
would generate a $b_Ls_L Z$ vertex.  New $SU(2)$ doublets ${U\choose D}_L$,
${U\choose D}_R$ mixed with the $R$-handed $d$-type quarks
would give rise to the new FCNC operator $b_R s_R Z$,
which is absent in the SM.
Both effects could appear simultaneously in the presence of
a set of multiplets of mirror fermions ${U\choose D}_R$, $U_L$, $D_L$.

It is easy to see why such flavor changing vertices are generated.
For each $L,R$ chirality state, the vector
$\Psi^o_{L,R}={d^o\choose D^o}_{L,R}$ of the ordinary $d^o$ and new
exotic $D^o$ quarks couples to the $Z$ through
the matrix of the isospin charges
\beq
T^3_L = -{1\over 2}\pmatrix{1&0\cr0&0\cr}, \qquad
T^3_R = -{1\over 2}\pmatrix{0&0\cr0&1\cr},
\eeq
which, in contrast to the SM, are not proportional to the identity matrix.
Then the isospin part of the neutral current for the mass eigenstates,
$\Psi_{L,R}=U^\dagger_{L,R}\,\Psi^{o}_{L,R}$, which
contains the flavor changing part of the coupling, reads
\beq \label{mixcurrent}
{\cal L}_{\rm mix} = {g\over 2\cos\theta_W}\, \sum_{i\neq j} \, \left[
\bar\Psi_{L}^i\, \kappa_{L}^{ij}\, \gamma^\mu\, \Psi_{L}^j +
\bar\Psi_{R}^i\, \kappa_{R}^{ij}\, \gamma^\mu\, \Psi_{R}^j \right] Z_\mu \,.
\eeq
Here and henceforth the hermitian conjugate is understood.
The strength of the flavor changing coupling $\kappa^{ij}_{L,R}$ is given by
\beq \label{kappa-FC}
-{1\over 2}\,\kappa_{L,R}^{ij} \equiv
  \left[ U^\dagger_{L,R}\, T^3_{L,R}\, U_{L,R} \right]_{ij},
  \qquad  (i\neq j) \,.
\eeq
As a result, a mixing of the $b$ with exotic $d$-type quarks will
induce a $bqZ$ ($q=d,s$) vertex at tree level.
We stress that a mixing is allowed only between states which carry
the same quantum number of an unbroken gauge group.
Therefore, for the corresponding photon coupling we have
$U^\dagger\,Q\,U \propto {\rm I}$ (with $Q=-\frac13\,{\rm I}$).
Namely, the electromagnetic couplings remain flavor diagonal.
The same holds for the part of the $Z$
coupling proportional to the electromagnetic generator.

The effective interaction (\ref{Lgeneral})
arising from (\ref{mixcurrent}) yields the coefficients
\beq \label{Cmix}
C_{L,R} = {g^2\over 4\cos^2\theta_W M^2_Z}\, \kappa_{L,R}^{bq}
\quad \Lra \quad \ct^2_{L,R} =
{3\over 4} \,  \sum_{q=d,s}|\kappa_{L,R}^{bq}|^2\,,
\eeq
where the factor of $3$ arises from the
sum over the neutrino flavors.
The current bounds on $|\kappa_{L,R}^{bq}|$ are
obtained from the limit on \bsmm\ (\ref{bsnnUA1}) (for $q=s$) and
from $B_d - \bar B_d$  mixing (for $q=d$)\cite{NirSilv,barger}.
Numerically, they are comparable with
the bound implied by the  limit (\ref{newlimit})
\beq
|\kappa_{L,R}^{bq}|  < 2.0 \times 10^{-3} \,.
\eeq

In some mass matrix models, the flavor changing couplings generated by the
ordinary-exotic fermion mixing are expected
to be of the order of the ratio of the light to heavy mass scales \cite{fit}
(see however \cite{nrt-new} for some interesting exceptions).
We see that the limits on $bsZ$ transitions are sensitive
to $b$ mixing with heavy particles up to a mass scale
$M\sim{\cal O}(\sqrt{m_b\,m_s}/\kappa^{bs})$, namely about $500\,$GeV.

\subsubsection{Models with large \bszp\ vertex}

Quite in general, new fermions have to be present in models based on
extended gauge groups (rank $>$ 4), since they are needed to ensure
the absence of anomalies in the new gauge currents.
$E_6$ models are a well known example where new $SU(2)$ singlets
$D_L$ and $D_R$ are present, together with new neutral gauge bosons
corresponding to the two additional Cartan generators of the group.
We generically denote the new gauge bosons as $Z'$,
which are coupled to the fermions through charges $Q'$.
A mixing with new {\it ordinary} quarks ({\it i.e.}, with conventional
$SU(2)$ quantum number assignments) would not affect the couplings
of the $Z$. However, as the matrix of the $Q^\prime$ charges is in general not
proportional to the identity,
$\kappa_{L,R}^{\prime\, ij}=
( U^\dagger_\alpha\, Q^\prime_{L,R}\, U_\alpha)_{i\neq j}$ will not
vanish, thus inducing a \bszp\ vertex.

In spite of the $1/M^2_{Z^\prime}$ suppression, the $Z^\prime$ mediated FCNC
are expected to be as large  as the corresponding transitions
induced by $Z$ exchange
\cite{lfc}.  This is because while in general the flavor changing mixings
affecting the $Z$ couplings are suppressed as the ratio of the light to heavy
fermion masses, no analogous suppression is expected for the mixings between
fermions of the same isospin, which affect only the $Z^\prime$ couplings.  The
absence of the suppression in the mixing can compensate for the
$M_Z^2/M^2_{Z^\prime}$ suppression of the $Z^\prime$ relative to the $Z$
amplitude, implying that the coefficients describing the  $Z$  and  the
$Z^\prime$  flavor changing effective interactions can be comparable in size
\cite{lfc}.

The analysis of the NP effects on an  enhanced \bszp\ vertex
parallels quite closely that of the large $bsZ$ models.
However, in the former case the enhancement of the \bsnn\ decay mode
relative to \bsll\ is not as large as in the latter one.
Since $\nu_L$ and $\ell_L$ appear in an $SU(2)$ doublet,
they have the same $Q^\prime_L$ charge. Then the ratio of the rates
of the two processes is given by
\beq \label{zpratio}
{\BR(\bsnn) \over \BR(\bsmm)} \approx
{Q^{\prime\,2}_{e_L}+Q^{\prime\, 2}_{\mu_L} +
Q^{\prime\,2}_{\tau_L} \over Q^{\prime\, 2}_{\mu_L}+Q^{\prime\, 2}_{\mu_R}} \,.
\eeq
In most models, the $Z^\prime$ couples universally
to all three generations, and then the above ratio cannot be larger than 3.
(This upper bound corresponds to the case
when the right-handed leptons are almost decoupled
from the $Z^\prime$, namely $Q^\prime_R \sim 0$. This can happen
in $E_6$ models in which the $Z^\prime$ arises as a particular combination
of the two additional $U(1)^\prime$ generators.)
As a result, these models are significantly more constrained by the limit
on \bsmm\ (\ref{brinc}) than the models with a large $bsZ$ vertex.
The UA1 limit implies $\BR(\bsnn) \lesssim 1.8 \times 10^{-4}$,
which is only marginally within the reach of the expected LEP sensitivity
(\ref{estbr}).

The expressions for the $C_{L,R}$ coefficients are similar to (\ref{Cmix})
\beq \label{CPmixn}
C_{L,R} = {g^{\prime\, 2}\over M^2_{Z^\prime}}\,
  {\cal F}_{L,R}(Q^\prime)\, \kappa_{L,R}^{\prime\,bq}  \quad \Lra \quad
\ct^2_{L,R} = 3 \left[\, {2\,r_g \cos^2\theta_W}\,
  {M_Z^2\over M^2_{Z^\prime}}\, {\cal F}_{L,R}(Q^\prime) \right]^2
  \sum_{q=d,s} |\kappa_{L,R}^{\prime\,bq}|^2 \,.
\eeq
Here $\kappa_{L,R}^{\prime\,ij}$ describe the strengths of the FCNC
couplings of the $Z'$ (the analogs of (\ref{kappa-FC})), and we defined
\beq
{\cal F}_{L,R}(Q^\prime) = Q^\prime(\nu)\,
  [Q^\prime(d_{L,R})-Q^\prime(D_{L,R})] \,, \qquad
r_g = {g^{\prime\,2}\over g^2} \,.
\eeq
For any particular $Z^\prime$ model, ${\cal F}_{L,R}$ and $r_g$ are known.
Then the limit (\ref{newlimit}) implies constraints on the ratios
$\kappa_{L,R}^{\prime\,ij}/M^2_{Z^\prime}$, typically about a factor two
weaker than the order $10^{-3}$ limits from the present bound on \bsmm.

Thus, in this class of models the largest allowed effects can hardly enhance
the \bsnn\ decay rate up to the estimated LEP sensitivity (\ref{estbr}).
However, the sensitivity of the \bsnn\ decay rate to $Z^\prime$ effects
can be larger in a class of  unconventional $E_6$ models \cite{unconvE6}.
In these models, the different generations are embedded in three
fundamental {\bf 27} representations in a generation dependent way,
implying different $Q^\prime_L$ charges
for the left-handed leptons of the different families.
In such a scenario the muon can be weakly coupled to the $Z^\prime$
without implying the same for $\nu_\tau$ and $\nu_e$, and then
the ratio of the \bsnn\ and \bsmm\ decay rates can exceed the
previously derived limit of 3.
For the unconventional $E_6$ models of Ref.~\cite{unconvE6}
we find that the ratio in (\ref{zpratio}) can be as large as
5. In this case  the constraint from \bsmm\ (\ref{brinc}) still allows
$\BR(\bsnn)$ up to $2.9\times10^{-4}$.

\subsubsection{Large $tcZ$ effective coupling}

In several NP models, new sources of FCNC are naturally suppressed, as
they are related to ratios between the masses of the fermions involved
in the flavor changing transitions and some large mass scale.
Due to the large value of the top mass,
such a suppression might not be effective for flavor changing transitions
involving the top quark \cite{topFC,Hall-Weinberg},
and theoretically the presence of a large $tcZ$ vertex
is indeed an open possibility (see \cite{peccei,laura,tcz},
and references therein).
This can be the case in models
which predict  new dynamical interactions of the top
quark \cite{TopColor,X-third}, in MHDM without natural flavor
conservation \cite{MHDM}, or in the presence of mixing with new $Q=\frac23$
isosinglet quarks \cite{tcz,buchi,parada}.

To date, the couplings of the top quark have not
been directly measured, and a large $tcZ$ vertex is not
constrained by the measurement of \bsg, neither by $B^0-\bar B^0$ mixing.
However, an anomalous $tcZ$ coupling will give new contributions
to the effective $bsZ$ vertex \cite{peccei,tcz}.
Contributions to the \bsnn\ decay
would arise from diagrams involving loops of charged
$W^\pm$ bosons and of unphysical $\phi^\pm$ scalars, with an insertion
of the $tcZ$ coupling on the fermion line \cite{peccei,tcz}.
Since these diagrams are not CKM suppressed (they are proportional to
$V_{tb}^*\,V_{cs}$), there is no additional suppression
beyond the loop factor.

The case of a tree level $tcZ$ vertex induced by
a mixing with new $Q=\frac23$ isosinglets was analyzed in \cite{tcz}.
In this particular case the vertex arises at tree level. Therefore, after
the underlying theory is fully specified, the computation
of the new penguin diagrams can be performed in detail,
yielding a finite result \cite{tcz}.
It was found that the new contributions to the $bsZ$ effective
vertex are always smaller in absolute value
than  the SM contributions, and opposite in sign. As a result, they interfere
destructively thus lowering the expected rates for \bsll\ and \bsnn.
Thus, an anomalous $tcZ$ vertex induced by mixing with $Q=\frac23$
isosinglets cannot be constrained by an upper bound on the branching ratio.

If the $tcZ$ vertex is an effective one, and
the underlying theory is not specified,
the expression for the loop induced $bsZ$ vertex by itself
is formally divergent, and it has to be regulated.
We take the result of the computation of the flavor changing penguins
from \cite{tcz} and we substitute the  function
resulting from the finite loop integration of that specific case,
with a regulator $\log{(\Lambda^2/m^2_t)}$.
We further assume that new effects
dominate over the SM contributions. This yields for the induced
$bqZ$ vertex
\beq \label{tcZeff}
\Gamma_{\rm eff}^{bqZ} \simeq {g\over\cos\theta_W }\,
  \left[{\alpha\over 4 \pi\sin^2\theta_W}\,
  (V^*_{tb}\,\kappa_L^{tc}\,V_{cq}) \log{\Lambda^2\over m^2_t} \right] \,,
\eeq
where $\kappa_L^{tc}$ parameterizes the strength of the $tcZ$ coupling.
For the dimensionless coefficient defined in (\ref{ctLR}) we get
\beq
\ct^2_L = {3\over 4}\, \left({\alpha\over 4\pi\sin^2\theta_W}\,
|\kappa_L^{tc}|\, \log{\Lambda^2\over m^2_t} \right)^2\,
\sum_{q=d,s} |V_{tb}^*\,V_{cq}|^2\,.
\eeq
Assuming $\Lambda\sim1\,$TeV, $|V_{tb}|\sim|V_{cs}|\sim 1$,
$|V_{cd}|\simeq 0.22$, we get from (\ref{newlimit})
\beq \label{tclimit}
 |\kappa_L^{tc}| \, \log {\Lambda^2\over m^2_t} < 0.55
\qquad \Lra \qquad
 |\kappa_L^{tc}| \lesssim 0.16 \,,
\eeq
which is comparable with the limit given in \cite{peccei},
once the differences in the  normalization and  in
the overall coefficient  in (\ref{tcZeff}) are accounted for.

We have chosen to present our bound in the form (\ref{tclimit}) for
uniformity of notations, and to allow for comparison with \cite{peccei}.
However, not to mislead the reader, a remark is in order.
$\kappa_L^{tc}$ parameterizes an effective vertex, and thus if
the heavy physics decouples from low energy, this parameter  must be
proportional to inverse powers of the NP scale $\Lambda$. Therefore,
even if this is not apparent from the notation we used
in (\ref{tcZeff}) and from the one used in \cite{peccei},
in the $\Lambda\to\infty$ limit the $bsZ$ coupling
approaches its SM value, as it should, in order to account
for the decoupling of heavy physics.

\subsubsection{Fourth Generation}

Extensions of the SM including a fourth generation are still an open
possibility \cite{fourth-fam}, and there are theoretically motivated models in
which the additional neutrino is naturally heavy \cite{zurabme}, thus escaping
the LEP limit on the number of light neutrino species.  It has also been shown
that SUSY models with four generations are consistent with unification, and by
imposing this requirement some conditions on the $t^\prime$, $b^\prime$,
$\tau^\prime$, and $\nu^\prime$ spectrum have been derived\cite{4gunion}.  In
these models large CKM mixings with a heavy $t^\prime$ may enhance the $Z$
and photon penguins \cite{soni}.

The constraints implied by the CLEO measurement of \bsg\ were presented in
\cite{hewett4bsg}.  After constraining the new mixing angles by the limits on
deviations of the $3\times3$ CKM matrix from unitarity,
it was found that a new $t^\prime$ quark
in the mass range $200-400\,$GeV is still consistent with experimental data.
The possibility of having a measurable enhancement of the $Z$ mediated FCNC
decays,  while keeping \bsg\ close to the SM prediction, relies on the fact
that the effective $bs\gamma$ vertex has a logarithmic dependence on
$m_{t^\prime}$, while for $bsZ$ this dependence is quadratic.
The dependence of the \bsnn\ rate on $m_{t^\prime}$ is numerically similar
to that of \bsmm. The different contributions of the box diagrams tend to
slightly enhance the neutrino decay over the charged lepton mode.
We find that the constraint (\ref{brinc}) still allows
\beq
  \BR(\bsnn)<3.7 \times 10^{-4} \,.
\eeq

In the presence of a heavy $t^\prime$ quark, the SM coefficient (\ref{HeffSM})
is modified according to
\beq
\ct^2_L = \Big(\ct_L^{\rm SM}\Big)^2\,
  \left[ 1 + \frac{V_{t^\prime b}^*\,V_{t^\prime s}}{ V_{tb}^*\,V_{ts}}\,
  \frac{X_0(x_{t^\prime})}{X_0(x_t)}\right]^2 \,,
\eeq
where, for simplicity, we assumed $|V_{t^\prime s}| \gg |V_{t^\prime d}|$ and
$|V_{ts}| \gg |V_{td}|$.  From the limit on \bsnn\ (\ref{newlimit}) we get
the bound
\beq
\left|\, 1+\frac{V_{t^\prime b}^*\,V_{t^\prime s}}{ V_{tb}^*\,V_{ts}}\,
  \frac{X_0(x_{t^\prime})}{X_0(x_t)}\, \right| < 2.8 \,,
\eeq
which is numerically similar to the limit implied by \bsll.

\subsubsection{Anomalous $WWZ$ couplings}

The $SU(2)_L\times U(1)_Y$ gauge symmetry of the SM fixes the dimension-4
operators that describe vector-boson self-couplings.  This symmetry should be
respected by any low energy effective theory, independently of possible NP
at energies  above the electroweak scale.  NP may still signal
itself in low energy experiments through higher dimensional operators,
suppressed by inverse powers of the NP scale  $\Lambda$.  While
dimension-6 operators modify the $WW\gamma$ and $WWZ$ vertices identically,
this is no longer true for dimension-8, and  higher dimensional operators
\cite{WWZ}.  Thus, as long as dimension-6 operators dominate the possible
deviations from the SM, the experimental measurement of \bsg\ implies that
also the \bsz\ coupling must be close to its SM value.

In general, dimension-6 operators, which are suppressed by two inverse powers
of the NP scale, are expected to dominate over dimension-8
operators, suppressed by $\Lambda^{-4}$.  However, in any perturbative
underlying theory the dimension-6 operators can only arise from loop diagrams,
and therefore they have an additional loop suppression factor of $1/16\pi^2$.
In contrast, dimension-8 operators can also arise at tree level \cite{WWZ}.
Thus, if the scale of new physics is below about $2\,$TeV, then dimension-8
operators can dominate over the dimension-6 ones.  In this case, the
measurement of \bsg\ implies no direct constraints on the \bsz\ vertex.

The effects of anomalous $WWZ$ couplings on rare $K$ and $B$ decays were
studied in \cite{genrare,heklmm,baillie,DaVa}.  In general, seven coupling
constants parameterize the $WWZ$ interaction \cite{HPZK}. In low energy
processes, when the external momenta can be neglected, only $g_1^Z$ and $g_5^Z$
contribute \cite{heklmm,baillie}.
A recent CDF measurement \cite{WWZexp} implies that $\Delta g_1^Z=g_1^Z-1$
has to be small.  Therefore, for simplicity we assume $g_1^Z=1$, and we
study only the effects of $g_5^Z$.  The $Z$ penguin (\ref{eqX}) contribution
is modified according to \cite{heklmm,baillie,DaVa}\footnote{%
We use the sign convention of \cite{DaVa} for $g_5^Z$, which differs from
that of \cite{heklmm,baillie}. }
\beq
  X_0(x_t) \to X_0(x_t) + g_5^Z \cos_W^2 W(x_t) \,, \qquad
  W(x) = {3x (1-x+ x\,\ln x )\over 4(1-x)^2} \,.
\eeq
The strongest published bound on $g_5^Z$ is obtained from
the measurement of $K_L\to\mu^+\,\mu^-$ \cite{heklmm}.
As we have discussed, because of the large long-distance
contribution to this process, such a bound is not very reliable.
We prefer to quote the bound resulting from $\BR(\bsmm)$.
For $m_t = 180\,$GeV we found that the limit on this decay mode implies
$-8.3 < g_5^Z < 3.7$.
This constraint allows  for
$\BR(\bsnn)$  up to the value in Eq.~(\ref{bsnnUA1}).
Our bound (\ref{newlimit}) implies
\beq
-8.6 < g_5^Z < 4.1\,.
\eeq

\subsubsection{Extended Technicolor (ETC)}

FCNC processes impose very strong constraints on technicolor models.  Either
``traditional" ETC models with a minimal set of interactions necessary for
third family quark mass generation, or ETC models which incorporate a
techni-GIM mechanism, yield a \bsg\ rate at most slightly larger than the SM
rate \cite{ETC1}.  While at leading order the $bs\gamma$ coupling is not
affected, these classes of ETC models typically induce large flavor changing
$Z$ boson couplings of the form $\bar s_L\,\gamma_\mu\, b_L\,Z^\mu$.
Techni-GIM models yield an even larger enhancement of 4-fermion interactions of
the form $O_L$ in (\ref{OL-OR}).

It was subsequently noted \cite{ETC2} that ETC models with a techni-GIM
mechanism also predict the \bsmm\ decay rate about a factor of 30 above the SM,
violating the experimental bound (\ref{brinc}), unless significant 
cancellations occur between various contributions.  
However, ``traditional" ETC models predict only about a factor of 4 
enhancement of \bsll\ over the SM. Therefore, these models are not yet 
excluded, and yield a similar enhancement for the \bsnn\ decay rate, 
which could be within the reach of the expected LEP sensitivity (\ref{estbr}).

\subsection{Unconstrained models}

\subsubsection{Light leptoquarks}

Leptoquarks (LQ) couple directly  leptons to quarks.
Such particles appear  in several extensions of the SM.
A comprehensive analysis of the experimental constraints
on the LQ couplings has been given in \cite{davidson},
and is summarized  in Table~15 of this reference.
After summing over all the possible neutrino flavors in the final
state, it turns out that the existing limits on the LQ
Yukawa couplings allow for \BR(\bqnn) up to ${\cal O}(10\%)$.
For the relevant couplings, our new limit (\ref{newlimit})
imposes much stronger constraints than the existing ones.
Several types of LQ are possible, and we adopt here the notations of
\cite{davidson}. LQ can be scalar $(S)$ or vector $(V)$ particles,
and can belong to different $SU(2)_L$ representations.
The $SU(2)_L$ singlets, doublets, and triplets are labeled
with the lower index $0$, $1/2$, and $1$, respectively.

The following LQ can mediate the \bsnn\ decay:
\beq
S_0\,, \qquad  \tilde S_{1/2}\,, \qquad  S_1\,, \qquad V_{1/2}\,,
  \qquad  V_1\,.
\eeq
Written in components, the relevant scalar and vector terms in the interaction
Lagrangian are \cite{davidson}
\beqa \label{LeffLQ}
{\cal L}_{LQ} = &-&  \lambda^{S_0}_{iq}\, \bar q^c_L\, \nu_L^i\, S_0
  + \lambda^{\tilde S_{1/2}}_{iq}\,\bar q_R\,\nu_L^i\,\tilde S_{1/2}^\dagger
  - \lambda^{S_1}_{iq}\, \bar q^c_L\, \nu_L^i\, S_1^\dagger \nonumber \\
&+& \lambda^{V_{1/2}}_{iq}\, \bar q^c_R\gamma_\mu\, \nu_L^i\,
  V^{\mu\,\dagger}_{1/2} + \lambda^{V_{1}}_{iq}\, \bar q_L\gamma_\mu\,
  \nu_L^i\, V^{\mu\,\dagger}_{1} \,,
\eeqa
where $q=d,s,b$, and it is understood that only the charge $\frac13$
component of the $SU(2)_L$ multiplets appears in (\ref{LeffLQ}).
For simplicity, we assume that all the $\lambda_{iq}$ couplings are real.
Integrating out the LQ fields and Fiertz transforming,  yields for
the coefficients of the effective four-fermion interaction
(\ref{Lgeneral}) induced by (\ref{LeffLQ})
\beq \label{CLR-LQ}
C_{L,R}^{qij} = \eta_{LQ}\, {\lambda_{iq}\,\lambda_{j3} \over m_{LQ}^2},
  \qquad  (q=d,s) \,,
\eeq
where $\eta_{LQ}=1/2\ (1)$ for scalar (vector) LQ.
The coupling $C_L$ is generated through $S_0$, $S_1$ and $V_1$ exchange,
while $C_R$ appears from $\tilde{S}_{1/2}$ and $V_{1/2}$.
As different types of LQ can exist, both $C_L$ and
$C_R$ can be simultaneously present. If in addition
the LQ carry some generation index, cancellations between different
generations  of LQ are also possible, and then the limit
(\ref{newlimit}) constrains only the total LQ-mediated rate.
For simplicity, we will restrict
ourselves to the case when only one type of LQ
is present, and it does not carry any generation index.

For scalar LQ the limit (\ref{newlimit}) implies the following new bounds
\beq \label{S-LQ-limit}
\lambda_{iq}\, \lambda_{j3} < 1.1 \times 10^{-3}\,
  \left({m_{LQ}\over 100\, {\rm GeV}}\right)^2\,,
\eeq
while for vector LQ
\beq \label{V-LQ-limit}
\lambda_{iq}\, \lambda_{j3} < 5.7 \times 10^{-4}\,
  \left({m_{LQ}\over 100\, {\rm GeV}}\right)^2\,.
\eeq
These bounds are much stronger than the existing limits \cite{davidson}.

Other LQ couplings involving the light fermions of the first and second
generations are constrained by the existing experimental data
to be much smaller than the limits (\ref{S-LQ-limit}) and
(\ref{V-LQ-limit}) \cite{LQ}.
However, if we have to learn a lesson from the hierarchy in the fermion
Yukawa couplings, then it seems natural to expect a large
hierarchy in the LQ couplings to the different generations as well.
This is the case, for example, in models that explain
the quark and lepton mass hierarchies as originating
from horizontal symmetries. If LQ exist, in these models they
couple more strongly to the third generation fermions \cite{QSA}.
Since three third generation fields and only one from the second generation
participate in the process $\bsntnt$,
any improvement in the search for the \bsnn\ decay would represent
an important test of these models.

\subsubsection{SUSY with broken R-parity}

In SUSY models it is usually assumed that R-parity is a good symmetry.
However, this is not necessarily the case, and one can construct SUSY
models with broken R-parity. We concentrate on the
MSSM without R-parity \cite{RSUSY}.
Extra trilinear terms are allowed in the superpotential, and some of them
can give rise to a large enhancement of the \bqnn\ decay rate.
Denoting by $L^i_L$, $Q^i_L$ and $d^i_R$ the chiral superfields
containing respectively the left-handed lepton and quark doublets,
and the right-handed down-type quark singlets of the $i$-th
generation, these terms read
\beq  \label{Wrpb}
  W_{\Rbs}= \lambda^\prime_{ijk}\, L^i_L\, Q^j_L\, \bar d^k_R \,,
\eeq
where, for simplicity, we assume the $\lambda^\prime_{ijk}$ couplings
to be real.  Omitting terms involving the $u_L$ and $\ell_L$ fermions
which are not relevant for the tree level $b\to q \nu_i \bar \nu_j$
transition, the Yukawa interactions of the R-parity breaking Lagrangian
generated by (\ref{Wrpb}) are \cite{bgh}
\beq \label{Rpbr}
{\cal L}_{\Rbs} =
\lambda^\prime_{ijk}\, [\,\tilde d^j_L\, (\bar d^k_R\, \nu^i_L) +
\tilde d^{k\, *}_R\, (\bar \nu^{i\, c}_L\, d_L^j)\,] \,.
\eeq

The exchange of $\tilde{d_R}$ and $\tilde{d_L}$ squarks gives rise
to the effective four-fermion interaction (\ref{Lgeneral}) responsible for
$B\to X_q\nu_i\bar\nu_j$,  with the coefficients
\beq \label{CLR-Rp}
 C_{L}^{qij} =
\sum_k {\lambda^\prime_{iqk}\, \lambda^\prime_{j3k} \over 2\,
 m_{\tilde{d}^k_R}^2} \,, \qquad
 C_{R}^{qij} =
\sum_k {\lambda^\prime_{ikq}\, \lambda^\prime_{jk3} \over 2\,
 m_{\tilde{d}^k_L}^2} \,, \qquad (q=d,s) \,.
\eeq
We note that, in contrast to the LQ case, here both the $O_L$ and $O_R$
operators are necessarily  present, due to the simultaneous appearance of
the $\tilde{d_L}$ and $\tilde{d_R}$ scalar superpartners of the $d_{L,R}$
 quarks.
For simplicity, we neglect possible cancellations
among the different $\tilde{d^k}$ ($k=1,2,3$) exchange amplitudes.
Then the bound (\ref{newlimit}) implies the following limits
on the product of R-parity violating couplings
\beq \label{Rp-limit}
\lambda^\prime_{iqk}\, \lambda^\prime_{j3k} < 1.1 \times 10^{-3}
  \left({m_{\tilde{d}^k_R}\over 100\, {\rm GeV} }\right)^2,  \qquad
\lambda^\prime_{ikq}\, \lambda^\prime_{jk3} < 1.1 \times 10^{-3}
  \left({m_{\tilde{d}^k_L}\over 100\, {\rm GeV} }\right)^2.
\eeq

These limits represent the strongest constraints on the product of
couplings involving the third generation, such as
$\lambda^\prime_{3q3}\,\lambda^\prime_{333}$,
$\lambda^\prime_{33q}\,\lambda^\prime_{333}$.
As in the LQ models, these couplings are expected to be particularly large
in models that relate their size (relative to the couplings involving the
lighter generations) to the fermion mass hierarchy \cite{bgnn}.

\subsubsection{TopColor models}

TopColor models \cite{TopColor}  attempt to explain the large value of the top
mass through the dynamical formation of a $t\bar t$ condensate.  In these
models, the basic assumption is that new dynamics strong enough to form chiral
condensates is effective for the third generation, which therefore is treated
differently from the first two.  As a consequence, TopColor models have
peculiar implications for the phenomenology involving the third generation,
such as top and bottom production \cite{HiPa} and the $Z\to b\,\bar b$
decay rate \cite{HiZh}.
In particular, they are expected to yield large effects in FCNC
$B$ decays involving third generation leptons \cite{BurTopC,BBHK}.

Several models can be constructed along these lines,
and we concentrate on the one studied in \cite{BurTopC}.
In this model the gauge symmetry breaking structure is
\beq \label{breakpat}
SU(3)_1 \times U(1)_1 \times SU(3)_2  \times U(1)_2 \to
SU(3)_{\rm QCD} \times U(1)_Y \,.
\eeq
Here $SU(3)_1 \times U(1)_1$ couples only to the third generation
while $SU(3)_2 \times U(1)_2$ couples only to the first and second
generations.  The quantum numbers under these groups coincide with those
under the usual $SU(3)_{\rm QCD}\times U(1)_Y$.
The breaking into the SM group (\ref{breakpat}) is induced by a
$\langle t\bar t\rangle$ condensate, which is generated at the $1\,$TeV
scale when the $SU(3)_1$ coupling becomes strong.
The initial symmetry is larger than the electroweak gauge group,
and this implies the existence of new massive gauge bosons
corresponding to the additional broken generators:
a color octet $B^a_\mu$ (topgluons) and a singlet $Z^\prime_\mu$.
We concentrate on the $Z^\prime$ boson since it can mediate \bsnn\ decays.
The couplings of  $Z^\prime$ to the fermions are given by
\cite{BBHK}
\beq \label{L-TopC}
{\cal L}_{\rm TopC} = g_1\,  f_i(\theta^\prime)
  \left( {1\over6}\, \bar Q^i_L\, \gamma_\mu\, Q^i_L +
  {2\over3}\, \bar u^i_R\, \gamma_\mu\, u^i_R -
  {1\over3}\, \bar d^i_R\, \gamma_\mu\, d^i_R - {1\over2}\bar L^i_L\,
  \gamma_\mu\, L^i_L - \, \bar\ell^i_R\, \gamma_\mu\, \ell^i_R \right)
    Z^{\prime\,\mu} \,,
\eeq
with $Q_L=(u,d)_L$, $L_L=(\ell,\nu)_L$, $g_1 \simeq 0.35$ is the
$U(1)_Y$ coupling constant, and $i=1,2,3$ is a generation index.

The most important difference between this model and the usual $Z^\prime$
models arises from the $f_i(\theta^\prime)$  factor, which enhances the
strength of the third generation couplings with respect to the
first and second generations.
One has $f_{1,2}(\theta^\prime)=\tan\theta^\prime$ for the first
two generations and $f_3(\theta^\prime)=-\cttp$ for the third generation.
In general $\cttp\gg1$ is expected,  in order to ensure that the condensate
forms in the top direction \cite{TopColor}.
After integrating out  the $Z^\prime$ boson and rotating the $d_{L,R}$ quarks
into the mass basis by the unitary matrices $U_{L,R}$,
the interaction Lagrangian (\ref{L-TopC}) gives rise to the following
coefficients for the effective four-fermion interaction (\ref{Lgeneral})
\beqa
C_L^{qii} = {1\over 12}\, {g_1^2\over M_{Z^\prime}^2}\,
  \kappa_L^{\prime\,bq} f_i  \quad &\Lra& \quad
\ct^2_{L} = \left( {\sin^2\theta_W\over 6}\,
  {M_Z^2\over M^2_{Z^\prime}}\right)^2
  \sum_{\scriptstyle i=1,2,3 \atop\scriptstyle q=d,s} f_i^2 \,
  |\kappa_{L}^{\prime\,bq}|^2 \,, \nonumber \\*
C_R^{qii} = -{1\over 6}\, {g_1^2\over M_{Z^\prime}^2}\,
  \kappa_R^{\prime\,bq} f_i  \quad &\Lra& \quad
\ct^2_{R} = \left( {\sin^2\theta_W\over 3}\,
  {M_Z^2\over M^2_{Z^\prime}} \right)^2
  \sum_{\scriptstyle i=1,2,3 \atop\scriptstyle q=d,s} f_i^2 \,
  |\kappa_{R}^{\prime\,bq}|^2\,.
\eeqa
Here $\kappa^{\prime\,bq}=\sum_j(U^*_{bj}\,U_{jq})f_j
\approx - (U^*_{b3}\,U_{3q})\,\cttp$
gives the flavor changing $Z^\prime$ couplings to the quarks
(the $L$ and $R$ chirality labels for $\kappa_{L,R}^{\prime\,bq}$
and for $U_{L,R}$ are understood).

A too large value of $\cttp$ would lead to a spontaneous
breaking of the chiral symmetry for the tau lepton.
This implies the bound \cite{BBHK,Kominis}
\beq \label{ctbou}
  \cttps < {8\pi^2 \over g_1^2} \quad\Lra\quad | \cttp| \lesssim 25 \,.
\eeq
Inserting the values of the various hypercharges given in (\ref{L-TopC})
into Eq.~(\ref{zpratio}), we find
\beq \label{btcchar}
{\BR(\bsnn) \over \BR(\bsmm)} =
  {f_1^2+f_2^2+f_3^2 \over 5 f_2^2} \approx {\cttpf \over 5} \,,
\eeq
where we assumed that NP effects dominate the decay rates,
and thus we neglected the SM contribution.
As is apparent from (\ref{btcchar}),
due to their different dependence on $\theta^\prime$, a measurement of
both  \bsmm\ and  \bsnn\  would allow us to separately determine
the value of this parameter.
The current limit on \bsnn\
in this model is obtained by combining Eqs.~(\ref{ctbou}) and
(\ref{btcchar}) with the experimental upper bound
on $\bsmm$ (\ref{bsnnUA1}). The  \bsnn\ rate is constrained
rather weakly by this process,  which still allows for
 a branching ratio up to about $ 10\%$.
The bound (\ref{newlimit}) gives the new stringent limits
\beq
|\cttp\, \kappa_{L}^{\prime\,bq}| < 5.4
  \left({M_{Z^\prime}\over1\,{\rm TeV}}\right)^2 \,, \qquad
|\cttp\, \kappa_{R}^{\prime\,bq}| < 2.7
  \left({M_{Z^\prime}\over1\,{\rm TeV}}\right)^2 \,.
\eeq

\subsubsection{Horizontal gauge symmetries}

Attempts to explain the hierarchical pattern of fermion masses and
mixings by some underlying dynamical interaction, are often based on
broken horizontal gauge symmetries \cite{YadFiz,horizontal,zurabme}.
In the non-Abelian case, the fermions of different generations
are assigned to some irreducible representation of the horizontal
gauge group.  Flavor changing transitions occur, suppressed by
the masses of the heavy horizontal gauge bosons.

The fermion mass pattern is generated through the so-called
universal see-saw mechanism \cite{uniseesaw}, in which the fermion masses
are suppressed from their natural scale $G_F^{-1/2}$, to the observed
values by inverse powers of some large mass scale, with  no need to
fine tune the Yukawa couplings \cite{YadFiz,horizontal}.
Various mass scales are associated with different stages of the
horizontal gauge symmetry breaking in such a way that the heavier fermions
of the third generation couple to the lightest horizontal gauge bosons.
As a consequence, rare FCNC transitions of the third family are
``naturally" enhanced.

To illustrate the main features of these models, we consider as a simple
example a model based on the horizontal gauge symmetry group $SU(3)_H\times
U(1)_H$ \cite{YadFiz}.  The standard $q$ quarks are assigned to the $(3,1)_H$
representation of $SU(3)_H\times U(1)_H$, while new isosinglet $Q$ quarks are
in $(\bar 3,-1)_H$.  The breaking of the horizontal symmetry is achieved
through the non-vanishing VEVs $\xi_\alpha$, of a set of SM singlet horizontal
scalars belonging to the $(\bar 3,2)_H$ or $(6,2)_H$, which also give large
masses to the heavy isosinglet quarks.  The mass terms for standard quarks are
provided by the VEV of the SM Higgs, $v=(2\sqrt{2}G_F)^{-1/2}\simeq175\,$GeV,
and by an additional VEV $\eta$ of a real scalar isosinglet belonging to
$(1,0)_H$.  Then, for the fermion mass terms we have
\beq \label{H-mass}
{\cal L}_{\cal M} \sim  \bar Q_L^i\, {\cal M}(\xi)_{ij}\, Q_R^j
  + \bar q_L^i\, Q_R^i\, v + \bar Q_L^i\, q_R^i\, \eta \,,
\eeq
where $i,j=1,2,3$ are flavor indices. We note that due to the
assumed horizontal gauge symmetry,
the second and third terms in (\ref{H-mass}) involving the light fermions
are the same for the three generations.

By assumption $v,\eta \ll \xi_\alpha\,$, and thus the quark masses
are generated via a see-saw like mechanism, yielding
for the mass matrices $m_q \approx v\,\eta\,{\cal M}(\xi)^{-1}$.
All information on the quark mass hierarchies and mixings
are contained in ${\cal M}(\xi)$, and depend on the hierarchical
structure of the VEVs $\xi_\alpha$. It is assumed that
$\xi_{1} \gg \xi_{2} \gg \xi_{3}$ in order to reproduce
the generation  hierarchy pattern.
For example, together
with  additional smaller induced VEVs contributing to $\cal M$,
they can  be chosen
to form a  Fritzsch structure \cite{Fritzsch}, yielding
\beq \label{xi-vevs}
\xi_{3} : \xi_{2} : \xi_{1} \sim
1: \sqrt{m_t\over m_u} :  \sqrt{{m_t\over m_u}\,{m_c\over m_u}}
  \sim 1: 190 : 3300  \,.
\eeq

The horizontal VEVs hierarchy induces the breaking chain
\beq \label{Hchain}
SU(3)_H\times U(1)_H \overarrow{\xi_1}
  SU(2)_H\times U(1)_H^\prime \overarrow{\xi_2}
  U(1)_H^{\prime \prime} \overarrow{\xi_3} I \,.
\eeq
In the first stage, four flavor changing gauge bosons
carrying family ``charge" plus a combination of the
three neutral generators acquire a large mass of order $\xi_1$,
leaving a residual $SU(2)_H\times U(1)_H^\prime$ symmetry unbroken,
which acts only on the second and third generations.
The four ``charged" bosons couple the first family fermions directly
(namely, not through small mixing angles)
to the fermions of the second and third generations,
and in particular they induce $\Delta S=2$ effective operators.
The  requirement that these operators will not generate  unacceptably
large contributions to the $\bar K^0-K^0$ mass difference implies
for the scale of the first breaking \cite{YadFiz}
\beq  \label{xi1-limit}
  \xi_1 \gtrsim 3\times 10^{3}\, {\rm TeV} \,.
\eeq
In the second stage, two charged bosons and a second neutral combination
acquire a mass of order $\xi_2$. These bosons give direct contributions to
the transition $B\to X_s\,\nu_\tau\,\bar\nu_\mu$,
 and as a consequence our limit (\ref{newlimit})
implies a lower bound on $\xi_2$ (see Eq.~(\ref{H-limit}) below).
The remaining local symmetry $U(1)_H^{\prime \prime}$ that acts only on
the third generation is broken at a scale $\xi_3$ much smaller than
$\xi_1$ and $\xi_2$. For example, if we assume the hierarchy
(\ref{xi-vevs}), the bound (\ref{xi1-limit}) implies the rather weak
constraint $\xi_3 >0.9\,$TeV.

We see that this scenario has the natural implication of a new $X^0_\mu$
boson corresponding to $ U(1)_H^{\prime \prime}$ in (\ref{Hchain}),
coupled only to the third generation, while due to the hierarchy
(\ref{xi-vevs}) the effects of
the heavier horizontal bosons are much more suppressed.
Due to the mixing between the quark mass eigenstates,
$X^0_\mu$ will also mediate flavor changing transitions.
The neutral current interaction Lagrangian
describing these transitions is analogous to (\ref{mixcurrent}), and
the terms relevant to the process we are interested in are
\beq \label{Leff-H}
{\cal L}_H = {g_H} \left[ \kappa_L^{bq}\, \bar b_L\, \gamma^\mu\, q_L +
  \kappa_R^{bq}\, \bar b_R\, \gamma^\mu q_R  +
  \bar\nu_{\tau L}\, \gamma^\mu\, \nu_{\tau L} \right] X^0_\mu  \,,
\eeq
where $\kappa_{L,R} = {U_{L,R}}^\dagger P_3\, U_{L,R}$ with
$P_3=$diag(0,\,0,\,1), and the matrices
$U_{L,R}$ rotate the $d_{L,R}$-type quarks
mass eigenstates.
Eq.~(\ref{Leff-H}) yields the coefficients
\beq \label{CLR-H}
C_{L,R} = {1\over\xi_3^2}\, \kappa^{bq}_{L,R}  \quad \Lra \quad
  \ct^2_{L,R} = {v^4\over \xi_3^4}\, \sum_{q=d,s}|\kappa^{bq}_{L,R}|^2\,.
\eeq
If the FCNC mixings $\kappa^{bq}_{L,R}$ are very small,
then in spite of the larger mass suppression the direct transition
$B\to X_s\,\nu_\tau\,\bar\nu_\mu$ mediated by the ``charged"
$SU(2)_H$ bosons of mass $M_{X^\pm} = g_H\,\xi_2$
(see (\ref{Hchain})) can give contributions to \bsnn\ which are
competitive to those mediated by $X^0_\mu$.
The rate for the $X_\mu^\pm$ mediated decay
$B\to X_s\,\nu_\tau\,\bar\nu_\mu$ becomes as large as
the rate for $B\to X_s\,\nu_\tau\,\bar\nu_\tau$ mediated by
$X^0_\mu$ when $|\kappa^{bq}_{L,R}| \sim \xi^2_3/\xi^2_2$.
Due to the different final states, the two amplitudes
do not interfere,  and
the coefficients $C_{L,R}$ for the second case follow easily from
(\ref{CLR-H}) by the substitution
$\kappa_{L,R}^{bs}/\xi_3^2 \to 1/\xi_2^2$.
Our limit  (\ref{newlimit}) implies the two bounds
\beq \label{H-limit}
 |\kappa^{bq}_{L,R}| < 6 \times 10^{-4}
 \left({\xi_3\over 100\, {\rm GeV}}\right)^2,  \qquad
\xi_2 > 6\, {\rm TeV}\,.
\eeq

A different scenario which also predicts a new $U(1)$ bosons coupled only to
the third family fermions was presented in \cite{X-third}.  Similar to
TopColor models, in this scenario the large value of the top mass is explained
by a dynamical scheme which implies a large isospin breaking for the $t$ and
$b$ masses.  However, that does not feed back directly into the $W$ and $Z$
masses.  The $U(1)$ gauge boson coupled to the third generation
is the remnant of the breaking of some large non-Abelian semi-simple gauge
group which embeds the HyperColor interaction responsible for the formation of
the isospin breaking condensate.  The phenomenology of such a gauge boson,
including the consequences on LEP physics due to its mixing with the $Z$, have
been extensively studied \cite{allHoldom}.
Recently it was also suggested that the \bsnn\ decay rate could be largely
enhanced because of the new FCNC contributions, and could even approach the
rate of semileptonic $B$ decay \cite{X-bsnn}.  Such a mechanism for increasing
the $B$ width was proposed as a possible solution to the claimed discrepancy
between the observed experimental value of the semileptonic $B$ branching ratio
and the theoretical predictions.  The limit (\ref{newlimit}) obviously rules
out this possibility, as well as any other mechanism attempting to achieve a
sizable increase of the $B$ width through an enhancement of decay modes
associated with large missing energy.

The interaction Lagrangian for the $X_\mu$ boson of this model, written in
the fermion mass basis, is identical to the Lagrangian in Eq.~(\ref{Leff-H}).
However, there is now one additional constraint for the $X_\mu$ mass
\cite{X-bsnn}
\beq
  {g^2_H \over M^2_X} = {G_F\over 2\sqrt{2}} \,.
\eeq
Using $M_{X} = g_H\,\xi_3$,
it is then straightforward to derive from
(\ref{CLR-H}) the expression for the $\ct_{L,R}$ coefficients
\beq \label{Holdom-CLR}
  \ct^2_{L,R} = {1\over 64}\,\sum_{q=d,s} |\kappa_{L,R}^{bq}|^2 \,.
\eeq
The NP parameters $\kappa^{bq}_{L,R}$ are denoted by
$\lambda^{L}_{i3}$ and $-\lambda^{R}_{i3}$ ($i=1,2$) in Ref.~\cite{X-bsnn}.
It was speculated that
$|\kappa^{bs}_{L}|^2+|\kappa^{bs}_{R}|^2 \approx 30\,|V_{cb}|^2$
could account for the possible discrepancy in the $B$ semileptonic branching
ratio \cite{X-bsnn}.  Our limit (\ref{newlimit}) implies a bound two orders
of magnitude smaller,
$|\kappa^{bs}_{L}|^2+|\kappa^{bs}_{R}|^2 < 2 \times 10^{-4}$.
For $q=d$, the limit $|\kappa^{bd}_{L}-\kappa^{bd}_{R}|<2\times10^{-3}$
can be derived from $B_d-\bar B_d$ mixing \cite{X-bsnn}.
However, there is no similarly strong limit on the individual
$|\kappa^{bd}_{L,R}|$ parameters.  From (\ref{newlimit}) we obtain
\beq \label{Holdom-limit}
  |\kappa^{bq}_{L,R}| < 1.4 \times 10^{-2} \qquad (q=d,s) \,.
\eeq

\section{Summary and conclusions}

In this paper we discussed the derivation of the first bound on the inclusive
\bsnn\ decay rate, using the large missing energy tag in $B$ decays at LEP.
We studied in detail the theoretical ingredients needed to carry
out such an analysis, and we found that the overall theoretical uncertainty is
small.  Therefore, this decay mode is well suited to search for physics beyond
the SM.  We translated the ALEPH bound on the \btn\ branching ratio, which
resulted from a search for $B$ decays with large missing energy, into a limit
on the \bsnn\ branching ratio.  To derive a numerical limit, we had to make a
number of conservative and simplifying assumptions.  Thus, the resulting bound
is weaker than what a dedicated experimental analysis will be able to achieve.
Our conservative upper bound is
\beq
  \BR(\bsnn) < 3.9 \times 10^{-4} \,,
\eeq
which is less than one order of magnitude above the SM prediction.
We estimated that using the full LEP--I data sample, the LEP
collaborations may be able to set a limit of order $(1-2)\times10^{-4}$.
Due to the theoretical interest in the \bsnn\ decay mode, we think it is
important that the LEP collaborations will perform a dedicated analysis
of this process.

We studied a variety of new physics models.  After discussing the constraints
from existing experimental data, we divided the NP models into three classes,
according to increasing allowed values of the the \bsnn\ branching ratio.

To class (A) belong those models in which the \bsnn\ branching ratio is already
constrained to be below the expected sensitivity of the LEP experiments.

Class (B) contains the models which allow for an enhancement of $\BR(\bsnn)$ up
to values that will be observable at LEP.  These models are listed in
Table~\ref{sumtab1}, together with the maximal \bsnn\ branching ratio they
allow for, after the existing constraints are taken into account.  These models
naturally evade the constraints imposed by \bsg.  The limits on the relevant NP
parameters which we obtain from the bound on the \bsnn\ decay rate, are
numerically close to the limits provided by the bounds on the inclusive and
exclusive \bsll\ decays.  However, our bounds are more reliable, as the \bsnn\
decay is theoretically cleaner.  We expect that in the near future, new results
emerging from CLEO, CDF and (hopefully) from LEP can compete in further
constraining these models.

The most interesting models for our investigation belong to class (C). For
the models  in this class, the bound on the \bsnn\ branching ratio implies new
constraints that are not  matched by other existing experimental data.  A
generic feature of these models is that they yield a natural enhancement of the
FCNC processes involving third generation fermions, without conflicting with
other constraints.  We derived new limits on the couplings of the third
generation fermions in models with leptoquarks, and in supersymmetric models
with broken R-parity.  Our bounds also imply stringent constraints on models in
which new gauge bosons are coupled dominantly to the third generation, such as
TopColor models, models based on non-Abelian horizontal gauge symmetries, and
other models attempting to explain dynamically the large value of the top mass.
The new bounds on the parameters of these models, implied by the limit on
\bsnn, are summarized in Table~\ref{sumtab2}.

In future $B$ factories much larger data samples will become available.  While
-- to our knowledge -- no detailed study concerning the possibility of
measuring the \bsnn\ decay rate at $B$ factories has been carried out, we
hope that it will be possible to perform such a search.
A precise measurement of the \bsnn\ decay rate would provide a very reliable
means of directly determining the $|V_{ts}|$ element of the CKM matrix.  This
would allow for an important new test of the unitarity of the CKM matrix.  If
deviations from the SM predictions will be detected in rare $B$ decays, it will
also be important to measure as many decay modes as possible.  The pattern of
deviations from the SM predictions in different decay rates would help us to
distinguish between various possible NP scenarios.

\acknowledgements
We are grateful to Ian Tomalin of ALEPH
for very helpful correspondence about the experimental analysis.
We also thank Yossi Nir and Mark Wise for discussions and comments on
the manuscript.
ZL was supported in part by the U.S.\ Dept.\ of Energy under Grant no.\
DE-FG03-92-ER~40701.

{\tighten

}  

\begin{table}
  \begin{tabular}{c|c}
Model                 &  Allowed branching ratio
  \qquad\qquad\qquad\qquad\qquad\qquad \\ \hline
SM                    &  $0.5 \times 10^{-4} $
  \qquad\qquad\qquad\qquad\qquad\qquad \\ \hline
FCNC $Z$              &  $3.5 \times 10^{-4} $
  \qquad\qquad\qquad\qquad\qquad\qquad \\
FCNC $Z^\prime$       &  $1.8 \times 10^{-4} $
  \qquad\qquad\qquad\qquad\qquad\qquad \\
Unconventional $E_6$  &  $2.9 \times 10^{-4} $
  \qquad\qquad\qquad\qquad\qquad\qquad  \\
Anomalous $tcZ$ vertex    &  $3.5 \times 10^{-4} $
  \qquad\qquad\qquad\qquad\qquad\qquad \\
Fourth generation     &  $3.7 \times 10^{-4} $
  \qquad\qquad\qquad\qquad\qquad\qquad \\
Anomalous $WWZ$       &  $3.5 \times 10^{-4} $
  \qquad\qquad\qquad\qquad\qquad\qquad \\
ETC                   &  $3.5 \times 10^{-4} $
  \qquad\qquad\qquad\qquad\qquad\qquad \\
  \end{tabular}
\vskip 6pt
\caption[tbmodels]
{Summary of the models belonging to class (B), which allow an enhancement
of the \bsnn\ decay rate up to a level observable at LEP.
The standard model (SM) prediction for \BR(\bsnn) is given for comparison
as the first entry.
For each model listed in  the first column, the second column gives the
maximal \bsnn\ branching ratio allowed by  the existing experimental data. }
\label{sumtab1}
\end{table}

\begin{table}
  \begin{tabular}{c|c}
Model  &  New bounds \qquad\qquad\qquad \\ \hline \\[-12pt]
LQ: \ \ $S_0$, $\tilde{S}_{1/2}$, $S_1$   &
$\displaystyle \lambda_{iq}\, \lambda_{j3} < 1.1 \times 10^{-3}
  \left({m_{LQ}\over 100\, {\rm GeV}}\right)^2$  \qquad\qquad\qquad \\[6pt]
\ \ $\,$ $V_{1/2}$, $V_1$   &
$\displaystyle \lambda_{iq}\, \lambda_{j3} < 5.7 \times 10^{-4}
  \left({m_{LQ}\over 100\, {\rm GeV}}\right)^2$  \qquad\qquad\qquad\\[12pt]
SUSY without R-parity  &
$\displaystyle \lambda^\prime_{iqk}\, \lambda^\prime_{j3k} < 1.1 \times
  10^{-3} \left({m_{\tilde{d}^k_R}\over 100\, {\rm GeV} }\right)^2$
  \qquad\qquad\qquad \\[6pt]
&  $\displaystyle \lambda^\prime_{ikq}\, \lambda^\prime_{jk3} < 1.1
  \times 10^{-3} \left({m_{\tilde{d}^k_L}\over100\,{\rm GeV}}\right)^2$
  \qquad\qquad\qquad \\[12pt]
TopColor  &
$\displaystyle |\cttp\, \kappa_{L}^{\prime\,bq}| < 5.4
  \left({M_{Z^\prime}\over1\,{\rm TeV}}\right)^2$  \qquad\qquad\qquad \\[6pt]
&  $\displaystyle |\cttp\, \kappa_{R}^{\prime\,bq}| < 2.7
   \left({M_{Z^\prime}\over1\,{\rm TeV}}\right)^2$  \qquad\qquad\qquad \\[12pt]
Horizontal gauge symmetry  &
  $\displaystyle |\kappa^{bq}_{L,R}| < 6 \times 10^{-4}
  \left({\xi_3\over 100\, {\rm GeV}} \right)^2 $  \qquad\qquad\qquad \\[6pt]
 &  $\displaystyle \xi_2 > 6 \, {\rm TeV} $  \qquad\qquad\qquad \\[6pt]
Model of Ref.\cite{allHoldom,X-bsnn} &
$\displaystyle  |\kappa^{bq}_{L,R}| < 1.4 \times 10^{-2} $
  \qquad\qquad\qquad  \\[-12pt] &
\end{tabular}
\vskip 6pt
\caption[tbmodels]
{Summary of the models belonging to class (C), for which essentially
no constraints on the allowed \bsnn\ branching fraction existed to
date. For each model listed in the first column, the second column
gives the bounds on the relevant model parameters implied by the new
limit $\BR(\bsnn)<3.9 \times 10^{-4}$. The indices $i,j,k=e,\mu,\tau$
and $q=d,s$ correspond  respectively to the neutrino and quark flavors
in the final state.}
\label{sumtab2}
\end{table}

\end{document}


{\tighten

\title{First limit on inclusive \bsnn\ decay \\ 
  and constraints on new physics \\ ${\rm }$ \\ 
  (Erratum)}

\author{Yuval Grossman\,$^{a}$, Zoltan Ligeti\,$^{b}$ and 
Enrico Nardi\,$^{a}$}

\address{ \vbox{\vskip 0.truecm}
  $^a$Department of Particle Physics \\
  Weizmann Institute of Science, Rehovot 76100, Israel \\
\vbox{\vskip -0.2truecm}
  $^b$California Institute of Technology, Pasadena, CA 91125 \\ ${\rm }$}

\date{July 16, 1996}
\maketitle

} 
\bigskip

In a recent article \cite{bsnn} we presented the derivation of the first limit
on the inclusive \bsnn\ decay rate.  Our work stemmed from the observation
that a published ALEPH bound on $\BR(\btn)$ \cite{ALEPH}, inferred from the
absence of large missing energy events in $B$ decays, implies also a limit on
$\BR(\bsnn)$.  We estimated this limit by comparing the missing energy spectra
in the two decay modes.  Theoretically, \bsnn\ is a very clean process, which
is also sensitive to several possible sources of new physics \cite{bsnn}. 
Therefore, a dedicated experimental search is important.  The result of such
an analysis by the ALEPH Collaboration, using the full LEP--I data sample, is
to appear soon \cite{private}.

As a result of discussions with members of the ALEPH Collaboration
\cite{thanks}, we found two errors in our numerical results.  The corrected
analysis yields a limit weaker than our original result \cite{bsnn} by about a
factor of three.  The corrections included here are formally beyond the free
quark decay result at tree level (of order $\alpha_s$, $\Lambda_{\rm
QCD}/m_b$, and higher).  However, they affect the final limit significantly,
due to the specific details of the experimental analysis.  

First, due to a mistake in our Monte Carlo code, the energy of the $X_s$
hadronic final state in the $B$ rest frame was evaluated as a fraction of the
$b$ quark mass, rather than as a fraction of the $B$ meson mass.  This resulted
in a harder missing energy spectrum than the true one.  Correcting for this
weakens the limit by about 35\%.  

Second, we neglected the effects of the invariant mass distribution of the
final hadrons, which we (implicitly) assumed not to extend to high mass states.
However, the invariant mass of the $X_s$ system can be well above $1\,$GeV
\cite{Adametal}.  Because of the large boost into the LEP laboratory frame and
the high missing energy range ($>35\,$GeV)  used in the analysis \cite{ALEPH}
(which is large compared to the average $B$ meson energy at LEP), neglecting
this effect yielded a missing energy spectrum in the laboratory frame
considerably harder than the correct one.  

It is not straightforward to include properly  this second effect into the
analysis. However, an  estimate can be obtained  by approximating the invariant
mass spectrum of the $X_s$ system  with a Gaussian distribution with mean
($\mu$) and variance ($\sigma$) fitted to the averages $\langle M_X^2\rangle$
and $\langle M_X^4\rangle$, as given in \cite{Adametal}.  Such a fit yields
$\mu\simeq1.35\,$GeV and $\sigma\simeq0.6\,$GeV.  There is about a 10\%
uncertainty in these values, due to their dependences on the heavy quark
effective theory parameter $\bar\Lambda\>(\simeq m_B-m_b)$.  Treating the $X_s$
mass distribution as independent of the energy spectrum, and including a
theoretical uncertainty of about 15\% related to the fitted values of $\mu$ and
$\sigma$, we find that the original limit given in \cite{bsnn} is weakened by
about a factor of three.  While small values of $E_X$ favor small values of
$M_X$ (beyond the trivial kinematic constraint $E_X>M_X$), the
correlation between $E_X$ and $M_X$ only slightly improve the bound.

In conclusion, the bound on $\BR(\bsnn)$ given in the Abstract and in
Eqs.~(1.2) and (6.1) of \cite{bsnn} is weaker by about a factor of three, 
while the limits on the new physics parameters, collected in Table 2, 
are weaker by about a factor 1.8.  
Until the ALEPH bound on the \bsnn\ branching ratio will appear, the
limit $\BR(\bsnn)<1.3\times10^{-3}$ should be used, instead of that
quoted in \cite{bsnn}.